\documentclass[elsart1p,letter]{elsart}

\usepackage{epsfig}

\def\cf4      {CF${_4}$\ }
\def\headtail {``head-tail''}

\usepackage{lineno}

\begin{document}

\begin{frontmatter} 

\title{Charge amplification concepts for direction-sensitive dark matter detectors}

\author[MIT,LNS]{D.~Dujmic\corauthref{cor}}
\corauth[cor]{Corresponding author.} 
\ead{ddujmic@mit.edu}
\author[MIT,LNS,MKI]{P.~Fisher}
\author[MIT]{G.~Sciolla}
\author[BU]{S.~Ahlen}
\author[MIT]{V.~Dutta}
\author[MIT]{S.~Henderson}
\author[MIT]{A.~Kaboth}
\author[NLR]{G.~Kohse}
\author[NSE]{R.~Lanza}
\author[MIT]{J.~Monroe}
\author[BU]{A.~Roccaro}
\author[Brand]{N.~Skvorodnev}
\author[BU]{H.~Tomita}
\author[MIT]{R.~Vanderspek}
\author[Brand]{H.~Wellenstein}
\author[MIT]{R.~Yamamoto}

\address[BU]{Physics Department, Boston University, Boston, MA 02215}
\address[Brand]{Physics Department, Brandeis University,  Waltham, MA 02454}
\address[MIT]{Department of Physics, MIT, Cambridge, MA 02139}
\address[LNS]{Laboratory for Nuclear Science, MIT, Cambridge, MA 02139 }
\address[MKI]{MIT Kavli Institute for Astrophysics and Space Research, Cambridge, MA 02139}
\address[NLR]{Nuclear Reactor Laboratory, MIT,   Cambridge, MA 02139 }
\address[NSE]{Department of Nuclear Science and Engineering, MIT,   Cambridge, MA 02139 }

\begin{abstract}
Direction measurement of weakly interacting massive particles in time-projection chambers 
can provide definite evidence of their existence and help to determine their properties.
This article demonstrates several concepts for charge amplification in
time-projection chambers that can be used in direction-sensitive dark matter search experiments.
We demonstrate reconstruction of the \headtail effect for nuclear recoils above 100~keV,
and discuss the detector performance  in the context of  dark matter detection and scaling to large detector volumes.
\end{abstract}


\end{frontmatter}

%
%

\section{Introduction}
\label{sec::introduction}

Weakly interacting massive particles (WIMPs) are good candidates for constituents of a dark-matter halo around
our galaxy, and a target of several experimental searches (see e.g. \cite{DMreview} for review). 
An annual oscillation in the signal at the level of 1-2\%  was recently confirmed by DAMA collaboration~\cite{Bernabei:2008yi}. 
If the asymmetry is due to Earth's motion around  Sun then much larger diurnal asymmetry in 
the WIMP direction is expected~\cite{directionality}.
However, the DAMA and other experiments that report zero signal~\cite{si-exp, sd-exp} use liquid and solid detector materials
that allow for more compact detectors, but destroy information on the  WIMP direction
and the sense of direction  (\headtail). 
Therefore, directionality is particularly valuable, both for suppressing background 
and for confirming the DAMA result by correlating a candidate dark-matter signal with astrophysical phenomena.

Reconstruction of the WIMP direction can be accomplished with detectors using low-pressure gas as the target material.
Several groups, DRIFT~\cite{DRIFT}, NEWAGE~\cite{Miuchi:2007ga},
MIMAC~\cite{Santos:2007ga} and DMTPC~\cite{Dujmic:2007bd}
have reconstructed low-momentum recoils created in elastic neutron scattering
using low-pressure time-projection chambers for dark matter searches. 
The reconstruction of the sense of direction (\headtail) 
using the scintillation profile of a recoiling track has 
been only recently achieved~\cite{Dujmic:2007bd}.

The low density of  gaseous detectors and the small  WIMP cross section necessitate the use of large  
detector volumes with fine detector granularities. A significant improvement of the current experimental limits
may require a ton-scale detector. 
In the case of \cf4 as the detector material, one ton of gas occupies a volume of approximately 
$16\times16\times16$~m$^3$ at 50~Torr of pressure. 
A 50~keV fluorine recoil created in a WIMP collision travels 1.5~mm at 50~Torr of pressure.
Therefore, a multi-cubic meter detector with resolution of the order of hundreds of micrometers 
is needed for a directional dark matter experiment,
requiring further progress in detector technology in order to observe WIMPs.

\section{Detector designs}

In this paper we propose and demonstrate several designs for time-projection chambers that can be used in large-volume, directional 
dark matter search experiments with \headtail discrimination.
In a time projection chamber, a WIMP creates a nuclear recoil that makes electron-ion pairs as it
slows down in the detector gas. 
We use \cf4 gas that has good
charge multiplication and scintillation properties~\cite{CF4 ionization,Pansky:1994zh,Kaboth:2008mi}. 
Fluorine, in addition, has non-zero angular momentum, which 
allows probing for spin-dependent dark matter interactions.
The electron diffusion limits the maximum drift distance to about 25~cm~\cite{Miuchi:2007ga,Dujmic:2007bd}.
%
%
%
The initial ionization electrons drift in electric field toward the charge-amplification region, where in case of \cf4 ,
the threshold for charge multiplication is approximately 45~V/(cm$\cdot$Torr)~\cite{CF4 ionization}.
Traditionally, design choices for amplification regions have been based on the
multiwire proportional chamber (MWPC)~\cite{Charpak:1997kd}, where strong electric fields are created in the vicinity of 
thin wires. The pitch between wires, and therefore the spatial 
resolution in MWPC is limited to above 1-2~mm due to mechanical and electrostatic reasons.
Finer spatial resolution can be achieved with micropattern detectors (e.g. \cite{micropattern}), 
but the  size and the gain of detector modules are constrained by manufacturing limitations.

\begin{figure}[hb]
\center
\begin{tabular}{l}
 a) mesh-copper\\
\includegraphics[width=8cm]{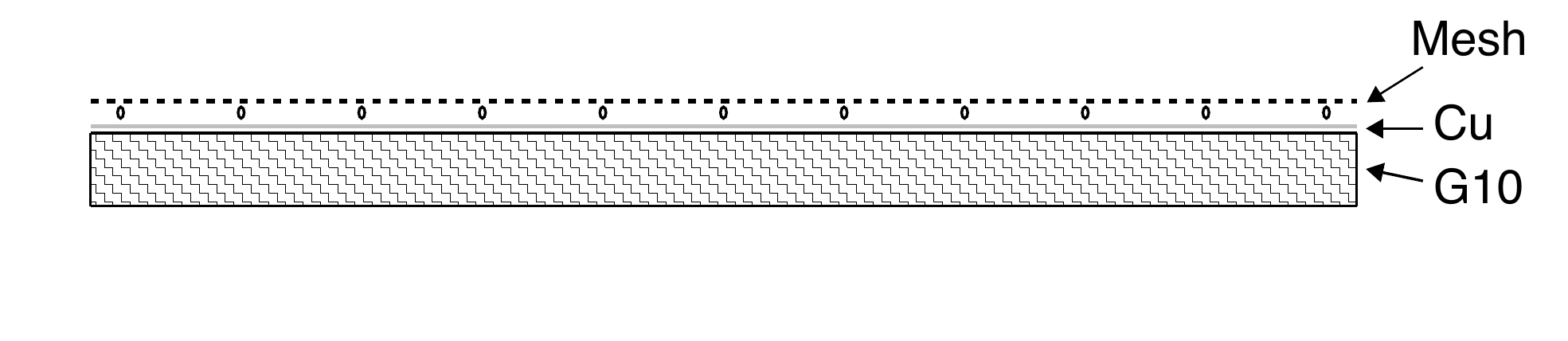} \\
 b) mesh-ITO film \\
\includegraphics[width=8cm]{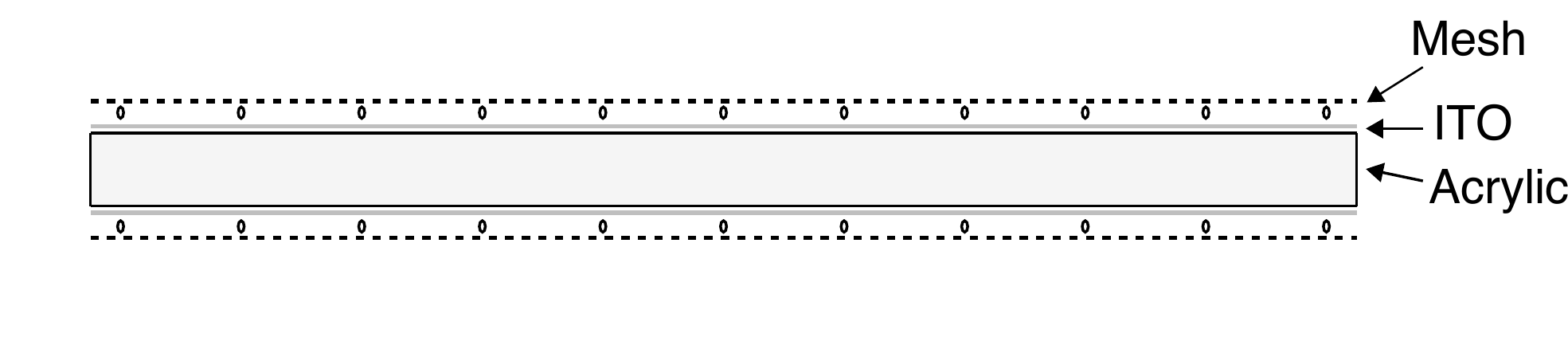} \\
 c) mesh-mesh\\
\includegraphics[width=8cm]{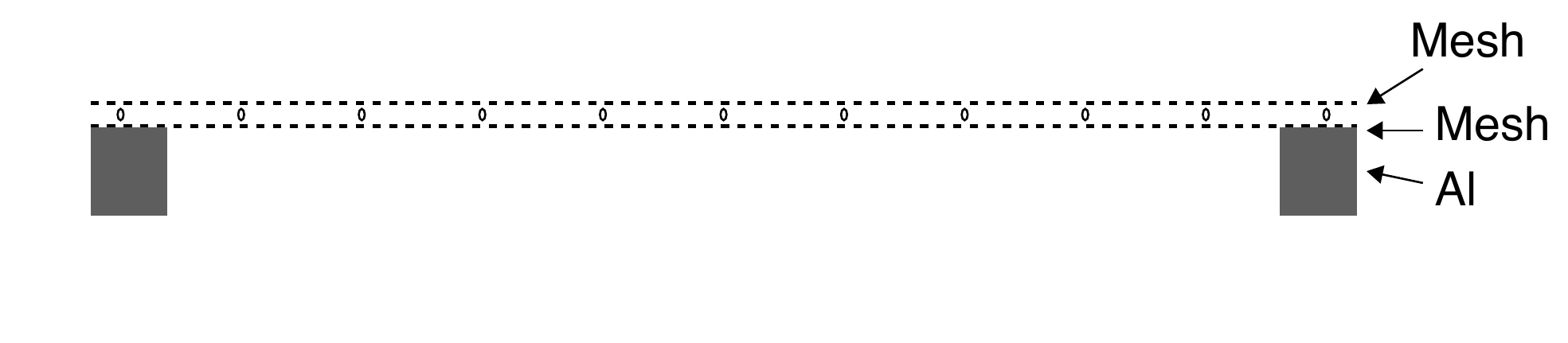} 
\end{tabular}
\caption{Schematics of amplification regions formed with 
a) mesh and a copper sheet,
b) mesh and ITO-film, and
c) two meshes. 
In all cases the electrodes are separated with 0.54~mm fluorocarbon wires.
\label{fg::amplifier}}
\end{figure}

In this paper we demonstrate  designs for micropattern detectors that provide good gain, fine granularity and scalability to
large volumes.
Electrodes used in the charge multiplication are made of woven meshes, indium-tin-oxide (ITO) films, or copper.
All designs allow production of large-area modules as the meshes and ITO foils are produced in 1.2~m-wide rolls.
In the first case shown in Figure~\ref{fg::amplifier}a), the amplification region is made of 
a stainless-steel mesh and copper-clad G10 board.
The stainless steel mesh is made of  28~$\mu$m diameter wires with periodicity of 256~$\mu$m, which gives optical transmittance of 77\%.
Woven meshes are used in air filtration systems and can be mass produced in a cost-effective way.
The mesh and the copper-clad board are separated by fluorocarbon resistive (fishing) wires of 0.54~mm diameter and spaced
every 2~cm, with transparency of 97\%. With this design, we demonstrate the reconstruction of 2D recoil segments with improved 
gain and no additional cost compared to our previous detector.
Further improvement in gain 
may be achievable with the use of transparent electrodes that allow scintillation light to be read out from two sides. 
We test a design shown in Figure~\ref{fg::amplifier}b), where the
amplification electrodes are made of ITO layers deposited on mylar foil and a stainless steel mesh.
ITO is 90:10 mix ratio, by weight, of indium-oxide (In$_2$O$_3$) and tin-oxide (SnO$_2$), respectively.
ITO is widely used in consumer electronics (e.g. touch screens, LCD screens) and available in
1.2~m wide rolls of  ITO-coated mylar foils. The  thickness of the ITO coating determines electrical and optical properties 
of the foil: we use a surface resistance of 15 Ohm/cm$^2$, which gives light transmittance of around 80\%.
In the third design shown in Figure~\ref{fg::amplifier}c), we use two meshes to create the amplification field
and the same fluorocarbon wires to separate the planes.
The transparency of electrodes can potentially increase the gain by almost a factor of two due to 
simultaneous readout of two drift volumes.

\begin{figure}[hb]
\center
\includegraphics[width=8cm]{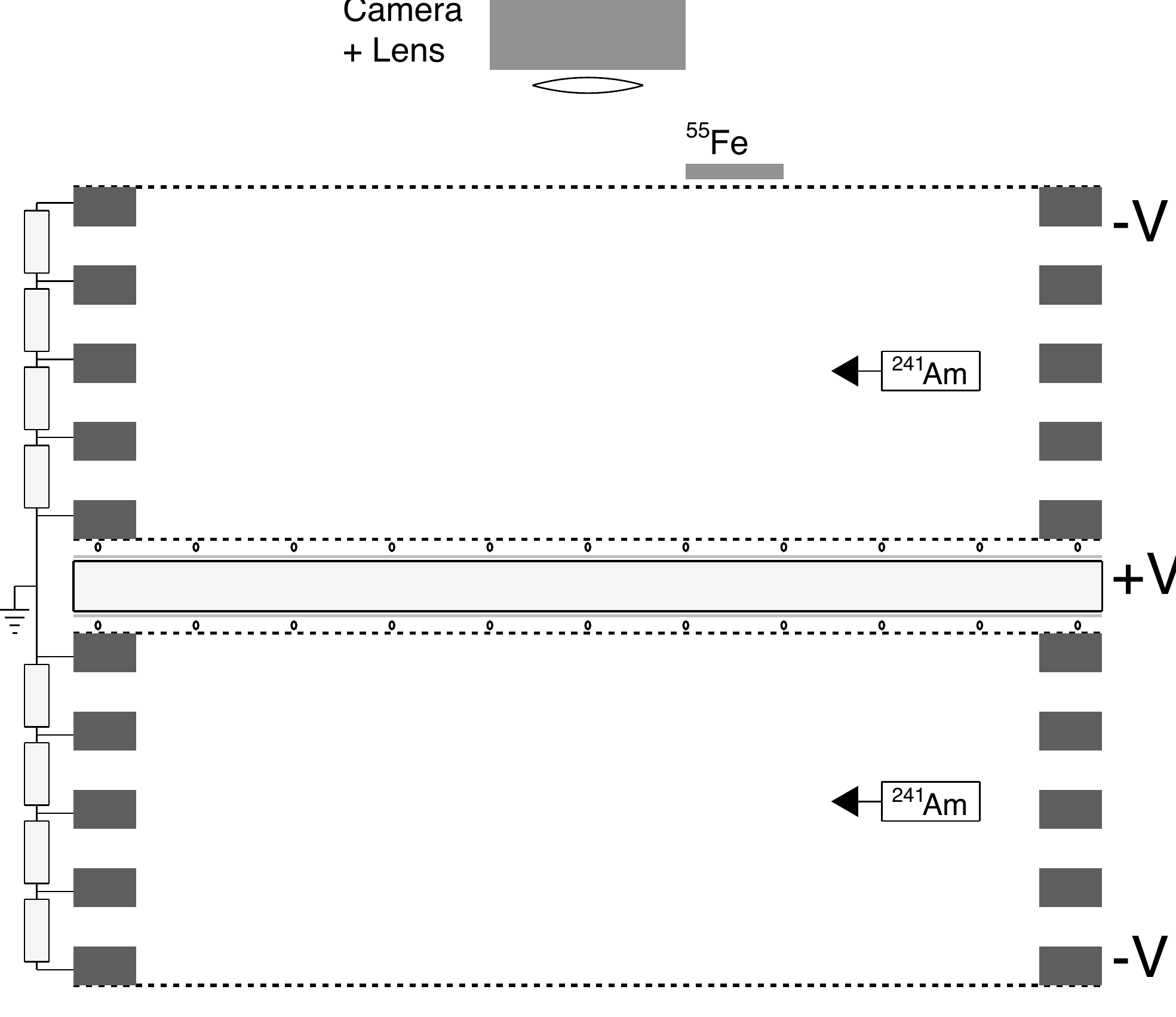} \\
\caption{A schematic of the TPC chamber used in these studies.
The inner diameter of the chamber is 23~cm, and the height of a drift region is 5~cm.
The ITO-mesh  configuration has two drift regions that can be read out simultaneously.
\label{fg::TPC}}
\end{figure}

The experimental setup with a time-projection chamber (TPC)  and a CCD camera for 
optical readout~\cite{ccd readout} of the amplification plane is shown in Figure~\ref{fg::TPC}.
The cathode mesh used to create the drift field has a periodicity of 312~$\mu$m and
wire thickness of 31~$\mu$m, which gives an optical transparency of 88\%.
The  drift distance is limited to approximatelly $\pm5~\rm{cm}$ by the size of the vacuum vessel.
The CCD camera is manufactured by Finger Lake Instrumentation and equipped with a Kodak KAF-0401ME chip with
a cooler that maintains the temperature in the range [$-20$,$-18$]~C to minimize  electronic noise (25~ADU). 
The photographic lens has an aperture ratio, f/\# of 0.95, and a focal length of 25~mm. 

%
%
%
%
%
%

\section{Results}
\label{sec::results}

\subsection{Detector gain}

We measure the detector gain using 5.5~MeV $\alpha$ particles from a collimated $^{241}$Am source.
The chamber is filled with  \cf4  at various pressures in the range 50-200~Torr.  
The drift field is 500~V/cm, and the amplification voltage ranges between  0.65 and 1.1~kV.
Images are taken sequentially with 500~ms exposure time and resolution of 96$\times$64 pixels, 
where each pixel is $488\times 488~\mu\rm{m}^{2}$ (72$\times$72~$\mu$m$^2$ on chip). 
Pixels that have intensity greater than 5 standard deviations from the mean dark field 
at least 10\% of the time are flagged as `hot channels' and excluded in the data analysis. 
ADC bias is corrected for by subtracting from each image the average of 100 images taken with the shutter closed. 

The gain of the detector is determined from the intensity of scintillation light recorded by the CCD camera 
in a 5~mm track segment close to the alpha source. The stopping power of alpha tracks in this region
is approximately uniform and we estimate it using the SRIM~\cite{srim} program, as  0.75~MeV/(mg/cm$^2$), which corresponds to
87, 130, 175, 264 and 355~keV of energy loss at 50, 75, 100, 150 and 200~Torr of \cf4 pressure, respectively.

We convert the observed light intensity into charge gain by accounting
for the transmittance of the amplification and drift meshes (70\%),
vessel window (90\%), camera lens and window (90\%). 
Scintillation light created in the lower drift region has an additional transmittance loss of approximately 60\%  as it 
passes through ITO layers and acrylic plate, or two amplification meshes.
We compute the lens acceptance to be $1.1 \cdot 10^{-3}$ and
estimate the average  CCD efficiency as $40\pm 10$\% using the manufacturer's  
quantum efficiency curve and the scintillation spectrum of \cf4 gas~\cite{Kaboth:2008mi}.
The gain of the camera is measured to be $1.6$ ADU/e$^-$.
The ratio of the total number of scintillation photons to the electrons in the avalanche is taken as 1/3~\cite{Pansky:1994zh, Kaboth:2008mi}.
We use $w=54$~eV for  the average ionization energy in \cf4 gas, or 18.5 ionization e$^-$ per keV of energy loss~\cite{Sharma:1998xw}.
Hence, the total gain, $g$ is computed from observed number of counts per keV of energy loss, 
$I_{CCD}$ as $g = I_{CCD}/2.5 \cdot 10^{-3}$.

\begin{figure}[p]
\center
\begin{tabular}{ll}
 a) mesh-copper\\
\includegraphics[width=9cm]{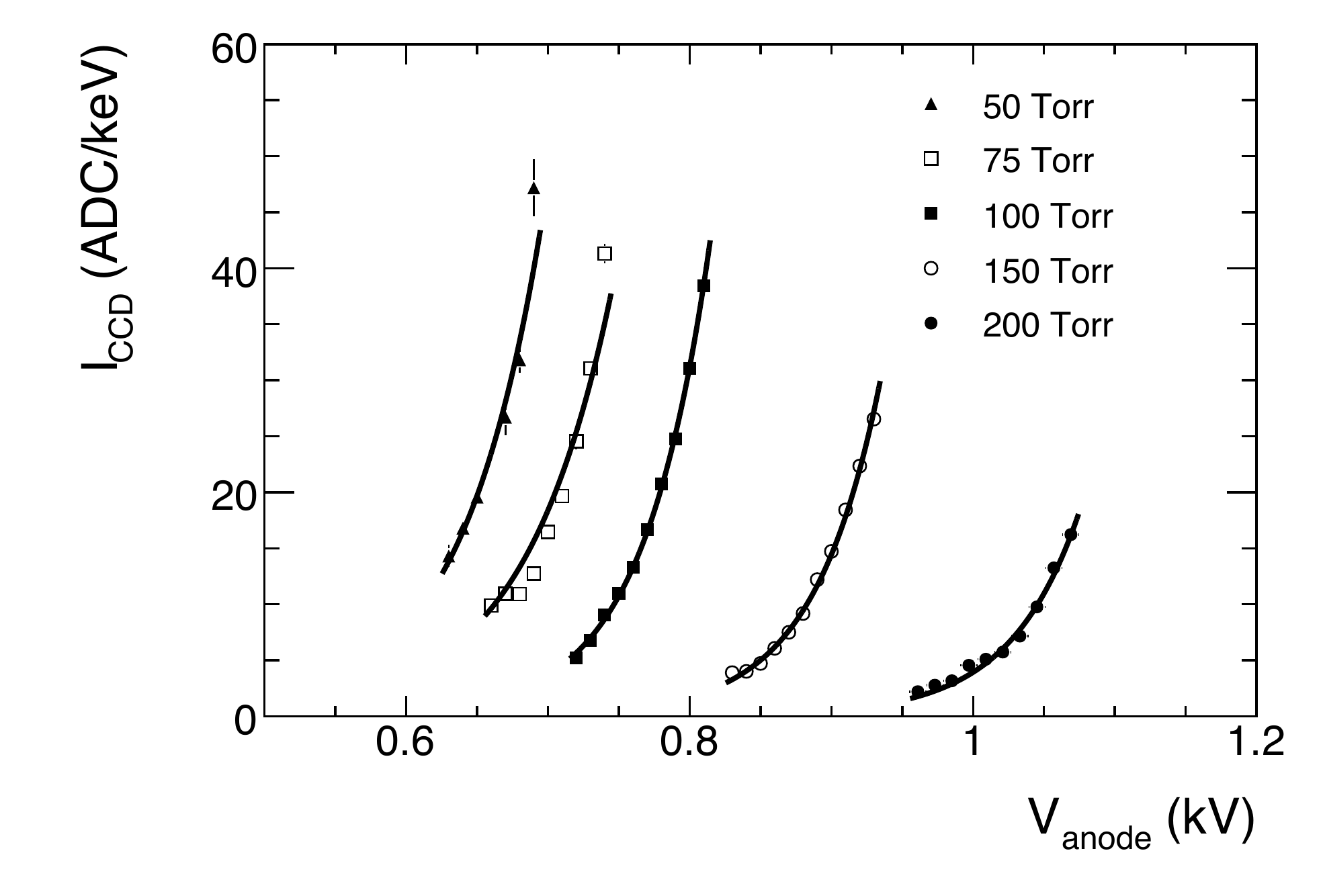} \\
 b) mesh-ITO film \\
\includegraphics[width=9cm]{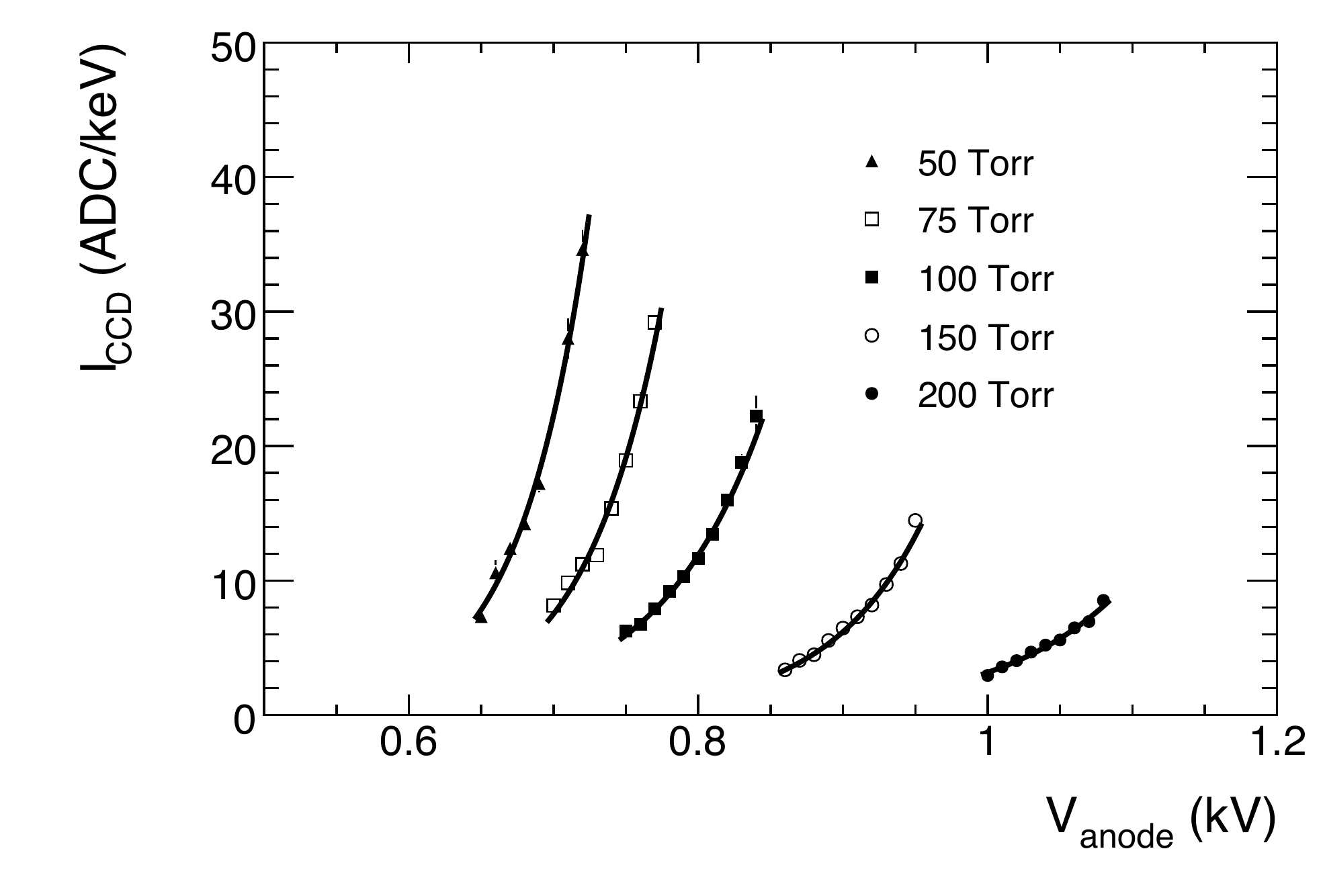} \\
 c) mesh-mesh\\
\includegraphics[width=9cm]{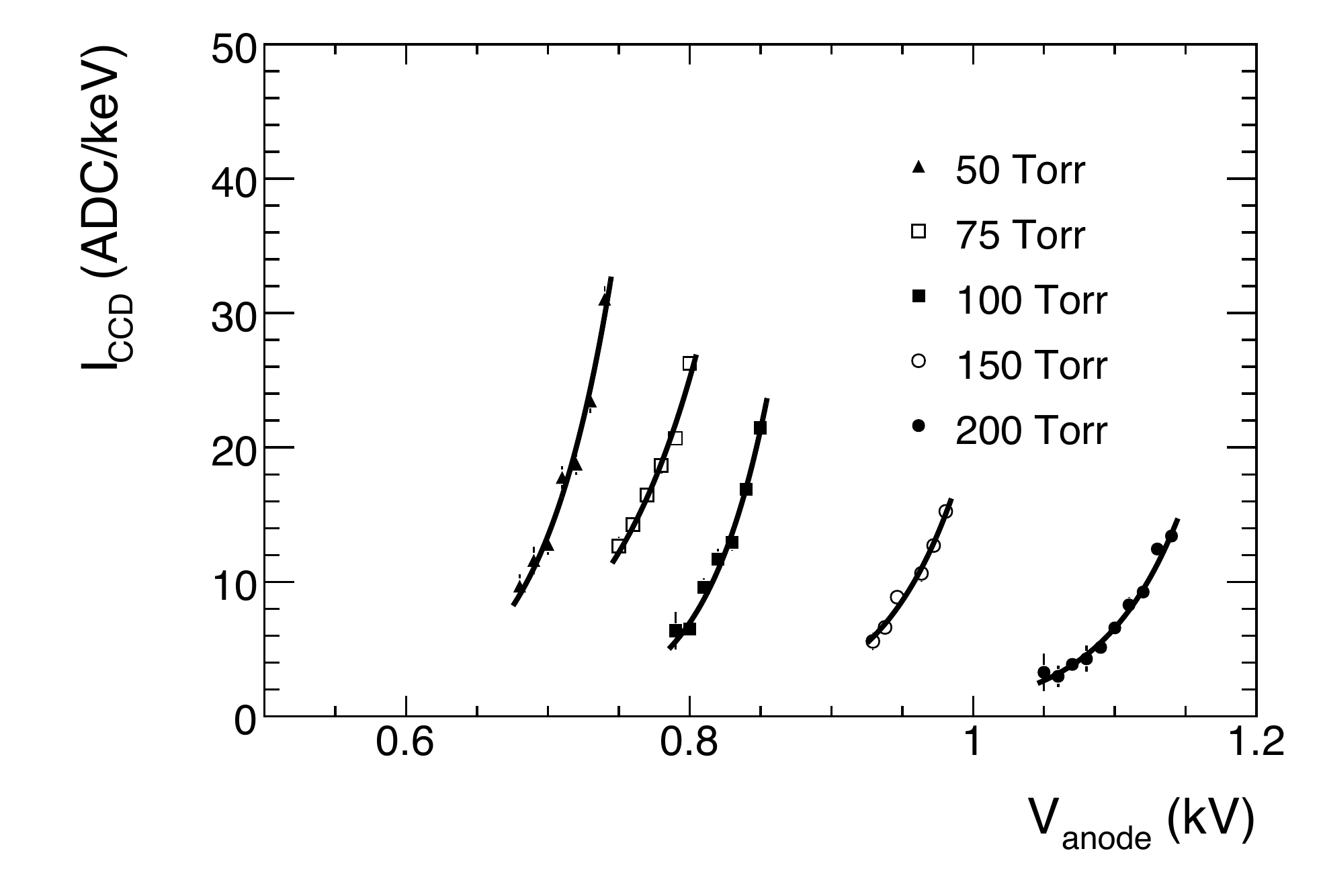} 
\end{tabular}
\caption{The gain measured in the three amplification designs. 
\label{fg::gain with alpha tracks}}
\end{figure}

We measure the gain in the mesh-copper detector (Figure~\ref{fg::amplifier}a) and 
plot the observed intensity $I_{CCD}$ as a function of the amplification voltage and pressure
in Figure~\ref{fg::gain with alpha tracks}a). The voltage is increased until the total charge created
in the amplification region reaches a sparking threshold.
The maximum gain varies from 15-45~ADU/keV, which is roughly 13 times larger than charge amplification achieved 
with wires~\cite{Dujmic:2007bd}, after accounting for different light collection efficiencies.
Since the lower ionization density allows  larger charge multiplication, the maximum gain increases with decreasing 
pressure, as is evident in Figure~\ref{fg::gain with alpha tracks}a).
The maximum electronic stopping power of alpha particles occurs in the Bragg peak at around 800~keV.
The same electronic stopping power is reached at 250~keV for fluorine ions,
presenting an upper-energy threshold for detection of nuclear recoils at a given gain.
We also compare different gaps between anode and ground planes using different widths of fishing lines.
After adjusting the amplification voltage, we find that the separation has no effect on the maximum achievable gain, 
which is limited by the total charge created in the amplification region.

\begin{figure}[hb]
\center
\includegraphics[width=10cm]{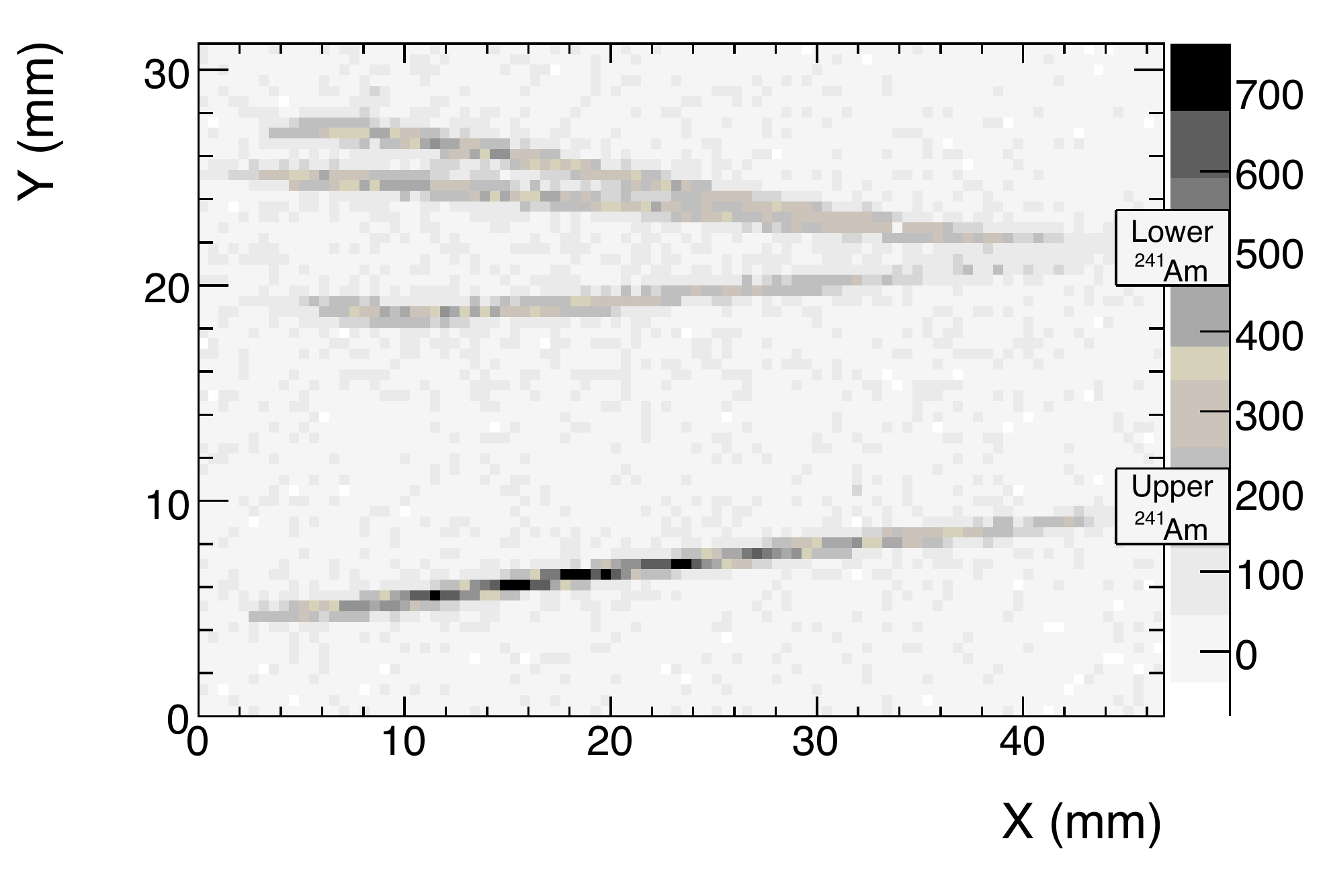} 
\caption{An image of alpha tracks from 500~ms exposure at 200~Torr with two $^{241}$Am sources placed in opposite
drift regions. The amplification electrodes are mesh and ITO foil.
\label{fg::images of alpha tracks}}
\end{figure}

The gain (or reduction in the number of CCD cameras) can be further improved by placing drift regions
opposite to each other along with transparent amplification electrodes, as shown 
in designs with ITO and all-mesh electrodes in Figures~\ref{fg::amplifier}b,c).
Two $^{241}$Am  sources are placed in opposite drift regions approximately 1~cm from the amplification region,
which is made of an ITO-mesh sandwich  (Figure~\ref{fg::amplifier}b).
Both drift regions are read out by the same CCD camera as shown in Figure~\ref{fg::TPC}.
An image taken with a 500 ms-exposure  of several alpha tracks  from two $^{241}$Am sources is shown in Figure~\ref{fg::images of alpha tracks}.
Note that the signal from tracks in the lower source is attenuated due to passage through two ITO-coated anodes and a ground mesh.
We measure the attenuation to be $(52 \pm 2)$\%, which is close to the expected attenuation of $(58 \pm 1)$\%
when we assume the same gain for the upper and lower amplification regions. 
We evaluate the gain using the signal from the upper source, and plot it in Figure~\ref{fg::gain with alpha tracks}b).
The gain is found to be 20-30\% lower than with the copper-mesh amplification plane. 
Further improvements are possible with foils that have higher resistivity (light transmittance) and better quality control
during shipment and module production, as our current foils arrived slightly damaged.

Finally, we test a detector that has the amplification plane made entirely of meshes.
The gain measurement is shown in Figure~\ref{fg::gain with alpha tracks}c).
Values are comparable with ITO-based detector and approximately 30\% lower than 
the copper-mesh detector. Operation with a readout of two drift regions can be realized
by adding a third mesh.

An energy resolution of approximately 10\%  and spatial resolution of  $400~\mu$m are 
measured here and found to be  comparable with our previous measurements using MWPC~\cite{Dujmic:2007bd}.
Contribution to the spatial resolution from the diffusion of electrons that are 2~cm 
above the amplification plane is  $140~\mu\rm{m}$, 
finite CCD bin size adds $488/\sqrt{12}\approx 140~\mu\rm{m}$,
and finite mesh pitch adds $256/\sqrt{12} \approx 74 \mu\rm{m}$.

\subsection{Gain uniformity}

The uniformity of the gain is demonstrated by observing a clear Bragg peak in mesh-ITO and mesh-copper detector
designs (Figure~\ref{fg::amplifier}a,b). A scintillation profile of an alpha track that has a range of approximately 4.5~cm in 250~Torr of \cf4
is shown in Figure~\ref{fg::images of alpha tracks}.

\begin{figure}[hb]
\center
\begin{tabular}{ll}
a) Moir\'{e} pattern & b) $^{55}$Fe exposure \\
 \includegraphics[width=6.5cm]{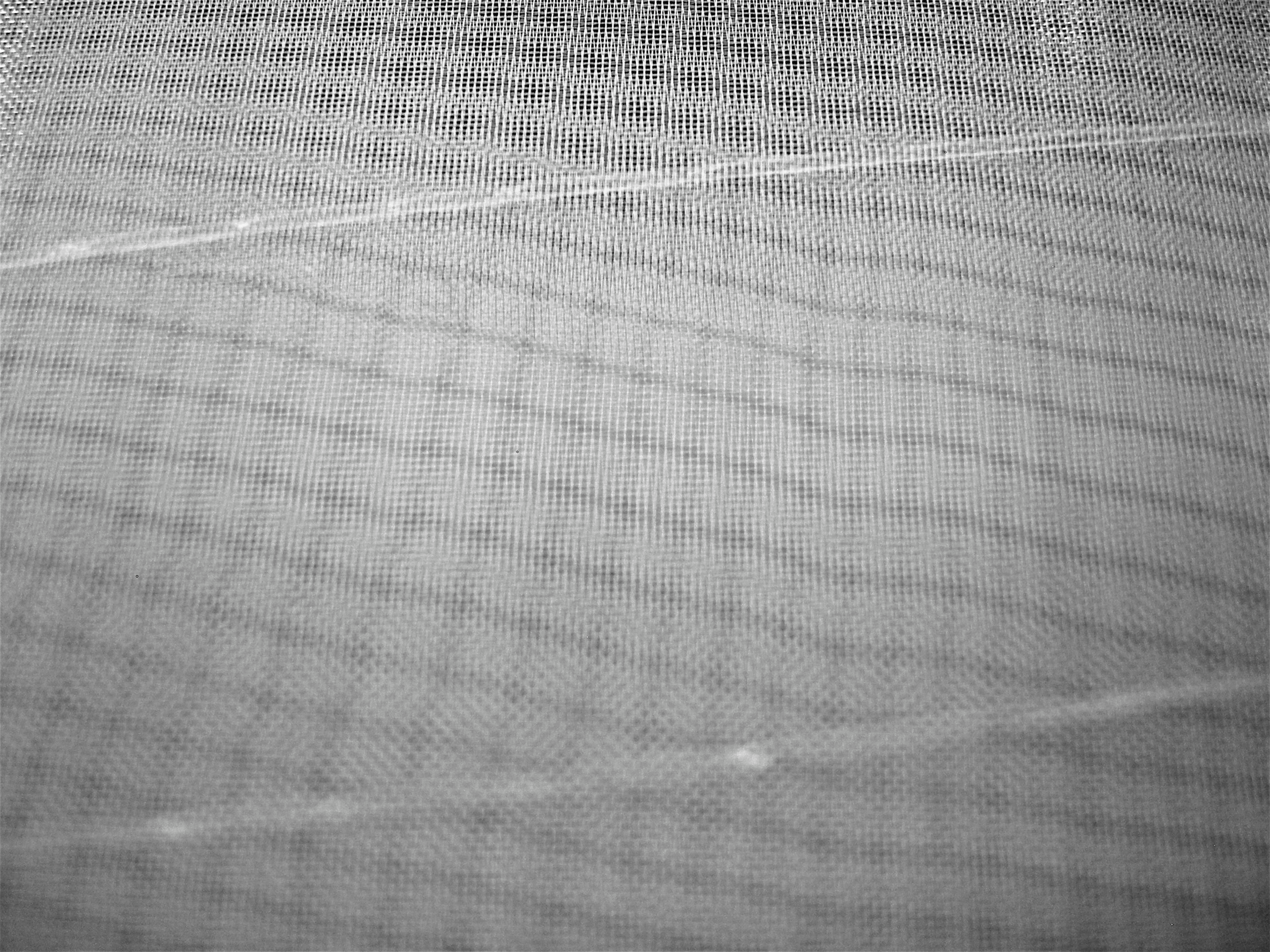} & 
\includegraphics[width=7.5cm]{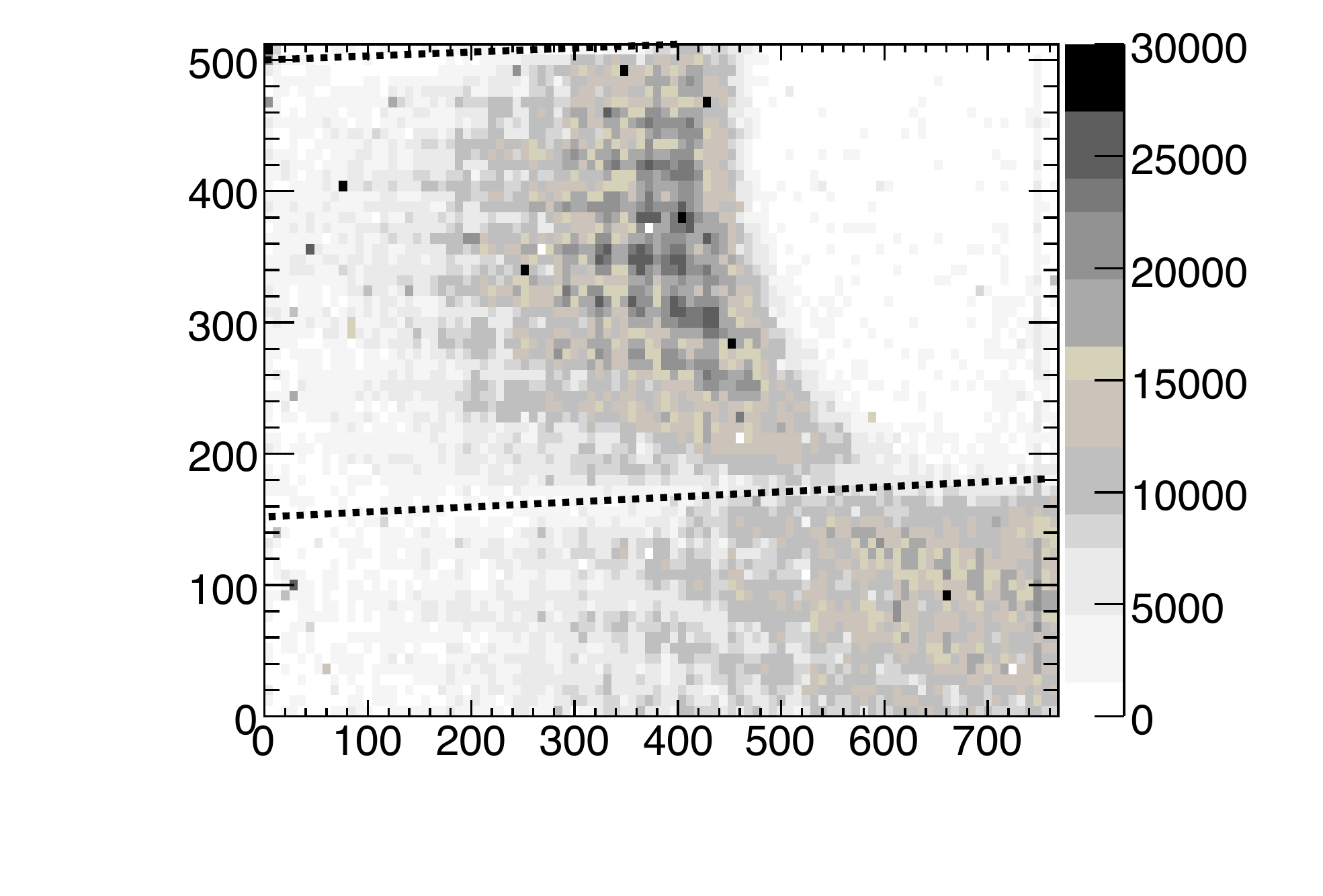}
\end{tabular}
\caption{Evaluation of gain uniformity in the mesh-mesh detector using
moir\'{e} pattern and $^{55}$Fe exposure. 
The distance between horizontal resistive wires is 2~cm.
\label{fg::gain uniformity of mesh-mesh detector}}
\end{figure}

In the case of the mesh-mesh detector (Figure~\ref{fg::amplifier}c), gain  non-uniformity can occur on a smaller scale if the relative alignment
between the meshes varies across the surface of the detector. The charge multiplication and light collection are more
efficient if wires of the anode mesh fall in between wires of the ground mesh, i.e. with offset of half of a mesh pitch
in both $x$ and $y$ directions.
The variation in relative alignment between the two meshes  may be the result of 
a non-zero relative angle, non-uniform mesh tensioning, 
or variation in spacing between the mesh planes.
We show an example of this  by taking a photograph  of light scattered from the bottom mesh and transmitted through 
the upper mesh, as shown in Figure~\ref{fg::gain uniformity of mesh-mesh detector}a). 
A moir\'{e} pattern with periodicity of roughly 2~mm is created by the change in relative alignment between two meshes. 
In order to evaluate the effect on the gain, we expose the detector to x-rays from an $^{55}$Fe with
gas pressure set to 400~Torr and anode voltage to 1.74~kV.  The average gain 
of approximately 9~ADU/keV is determined from the total light emitted by 5.5~MeV alpha particles.
An $^{55}$Fe source is  mounted on the top of the cathode mesh and several minutes of 1-second exposures are taken, with 
accumulated CCD image  shown in Figure~\ref{fg::gain uniformity of mesh-mesh detector}b).
The average light intensity decreases with distance from the source due to the reduced flux of x-rays.
The semi-circular shadow on the right is due to the Fe source, and the long shadows are due to fishing line separators.
The checkered light pattern seen in the plot is due to variation of the gain and 
follows the moir\'{e} pattern. 
Using pixels that have equal distance from the source, we 
estimate a gain variation to be approximately 50\% between maxima and minima.
The non-uniformity in gain can be minimized, if not avoided, by applying extra care during fabrication of detector.
Elimination of the moir\'{e} pattern during the mesh stretching and assembly 
can be used as a monitor for quality control. Any remaining gain variation can be accounted for by calibrating with x-ray sources.

\subsection{``Head-tail'' effect}

%
%
%
%
%
%
Finally, we demonstrate the ability to determine the directional sense of low-momentum nuclei, the \headtail effect.
The ionization rate by a low-momentum  nucleus decreases as it slows down in the detector gas,
and the sense of its direction can be deduced from the scintillation profile along the track.
We place a $^{252}$Cf source approximately 2~m from the detector's view-field of $4.8\times 3.2$~cm$^2$, 
resulting in a well determined direction of the neutron flux. 
We use the copper-mesh amplification plane (see Figure~\ref{fg::amplifier}a) with the anode voltage set to 740~V and 
the \cf4 pressure at 75~Torr.
Typical ranges for nuclear recoils created in neutron scattering are of the order of a few millimeters.

\begin{figure}[hb]
\center
\begin{tabular}{ll}
\includegraphics[width=6.5cm]{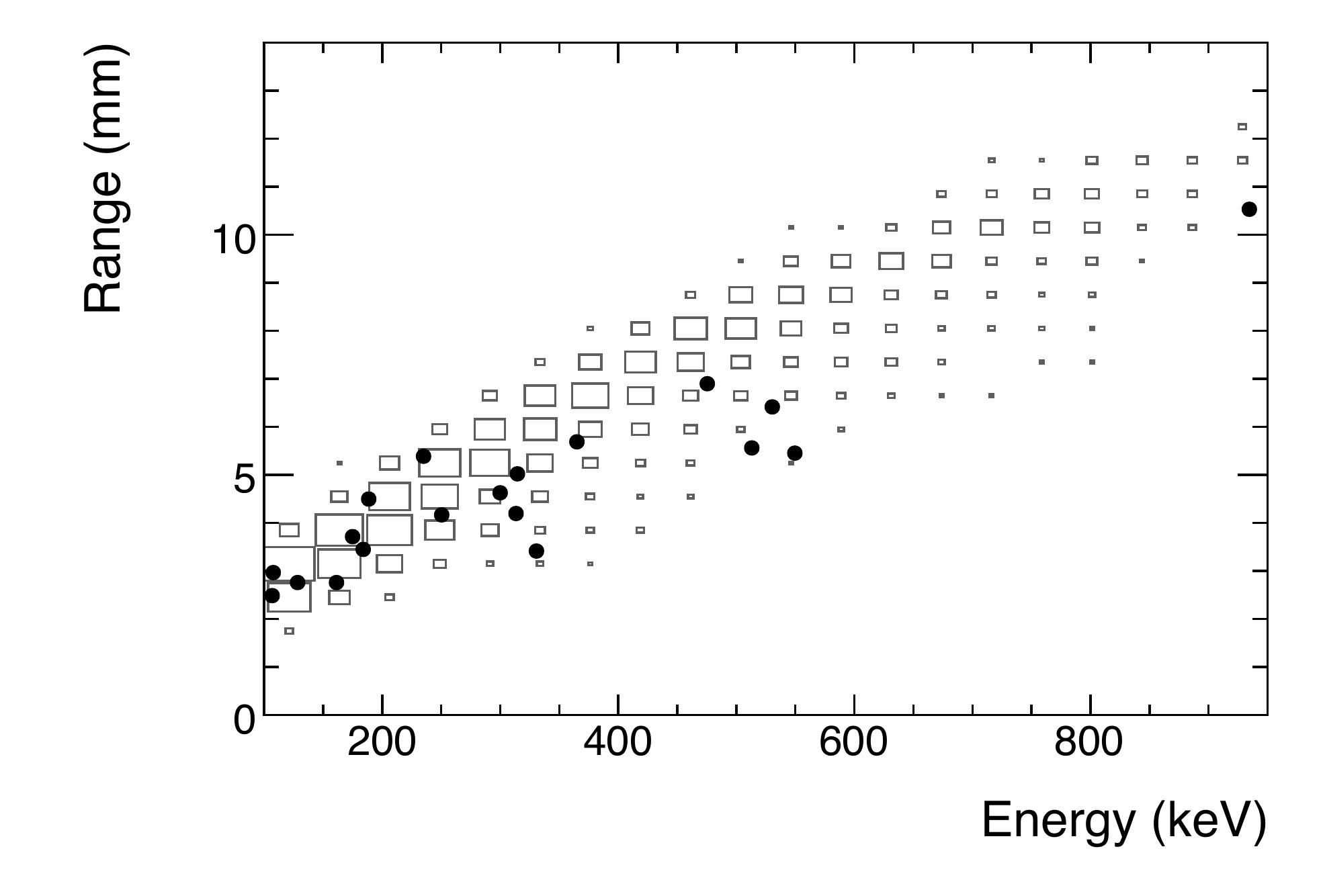} &
\includegraphics[width=6.5cm]{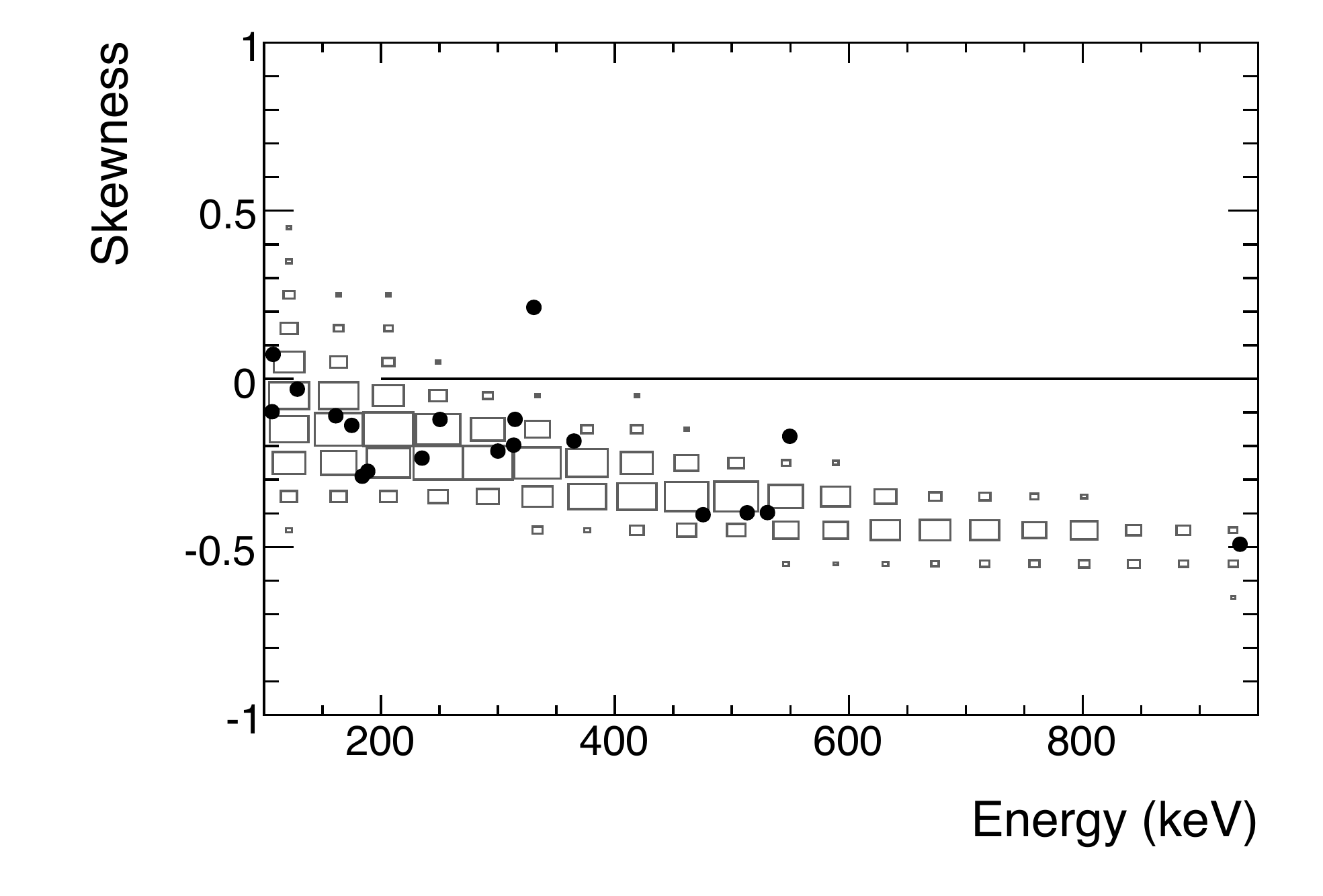} \\
\includegraphics[width=6.5cm]{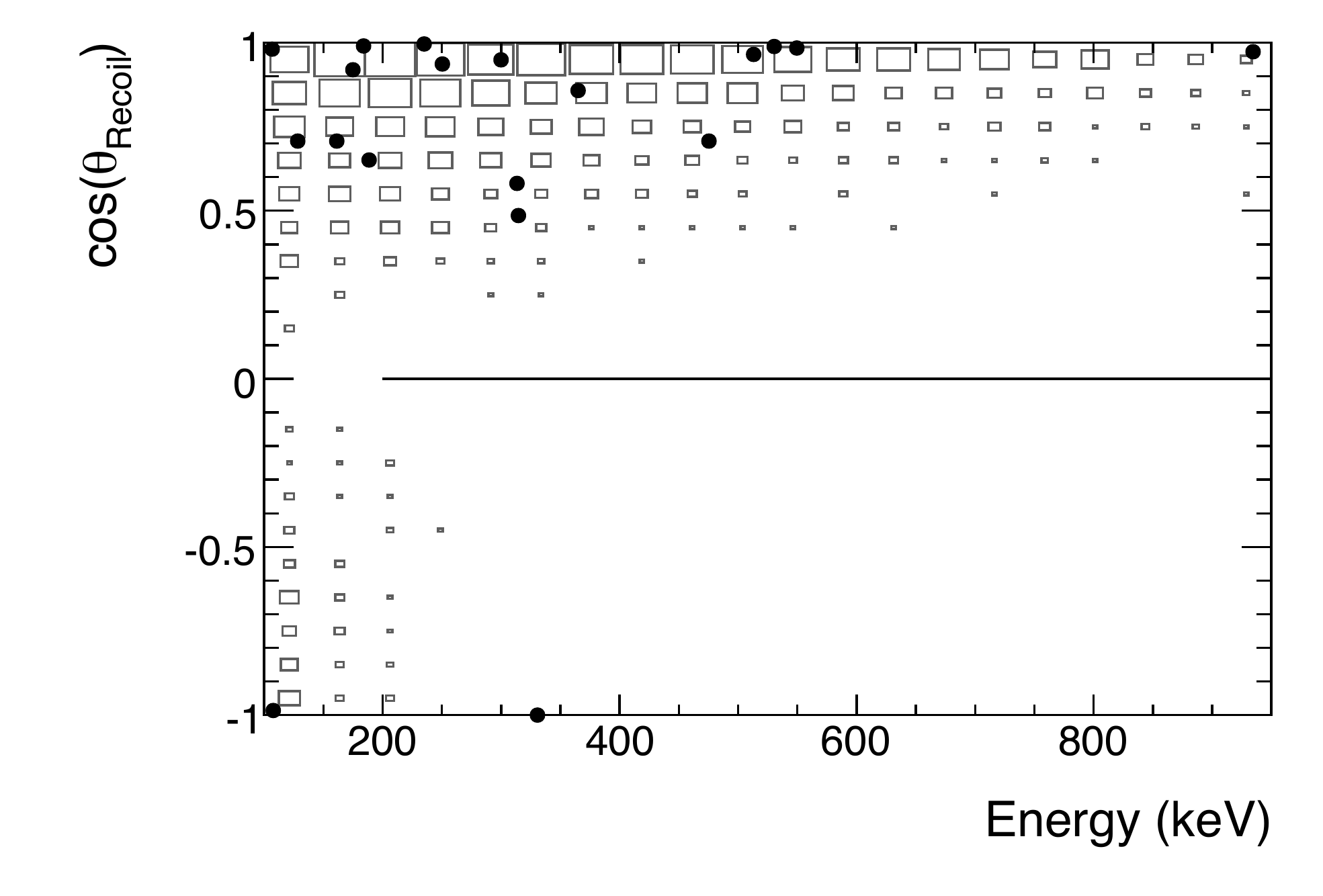} &
\includegraphics[width=6.5cm]{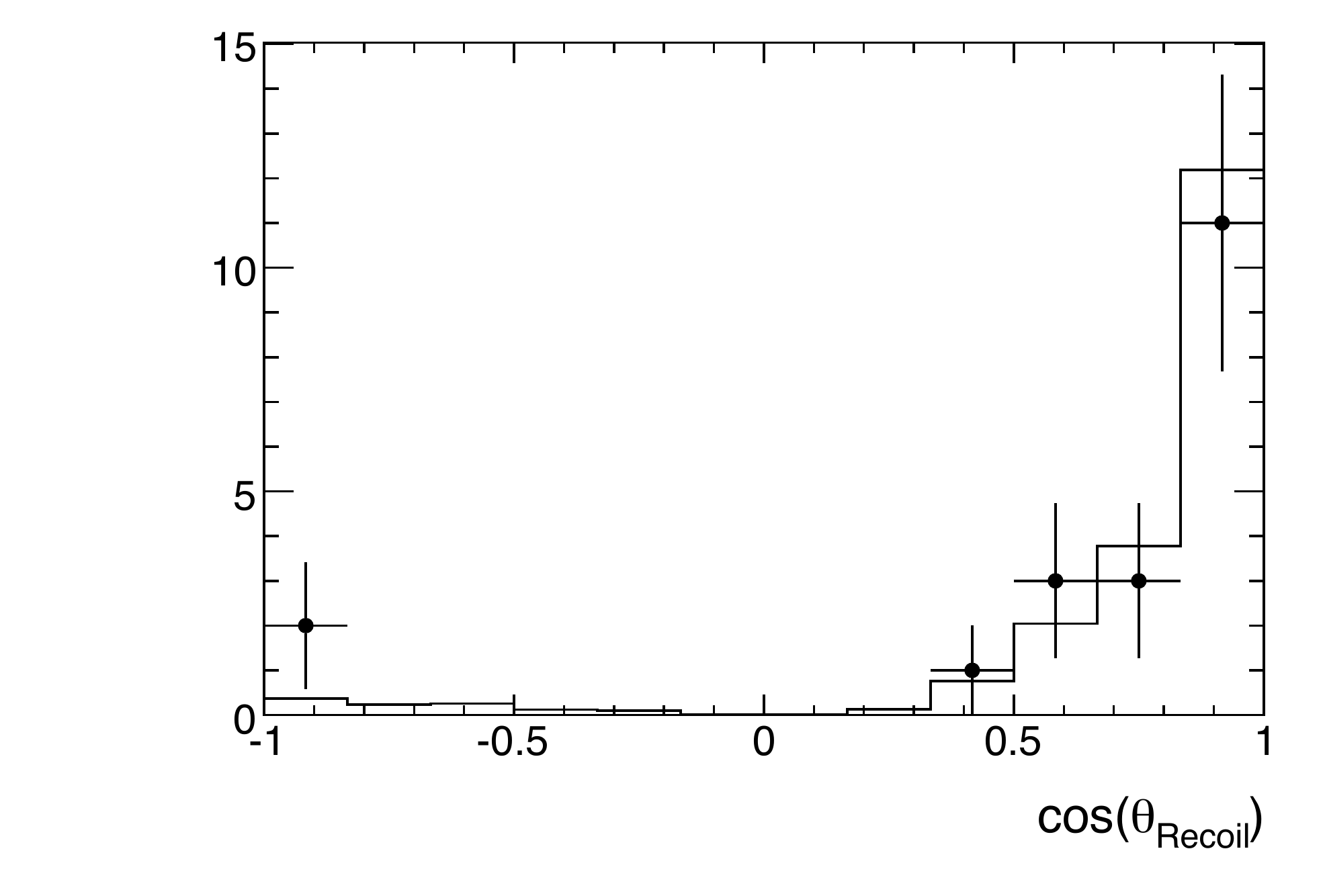} \\
\end{tabular}
\caption{Range (top left), skewness (top right), and cosine of 2D recoil angle (bottom left) vs. reconstructed energy 
for nuclear recoil candidates in a $^{252}$Cf exposure at 75~Torr. Black points are data, the box-histogram is simulation.
Signed distribution of the cosine of 2D recoil angle (bottom right), with data and simulation normalized to the same area.
\label{fg::energy vs range}}
\end{figure}

Taking 6000 1-second exposures without a trigger, we search for clusters of pixels that are 3 standard
deviations above the CCD noise. In most cases there is only one cluster per event, but otherwise we select the most
energetic cluster. The energy is computed from the sum of pixel yields after background subtraction, using 
calibration resulting from inserting an alpha source three times: before, in the middle and after
the neutron exposure. We find an average gain of 27~ADU/keV. The gain  decreases by 8\% during the measurements, 
probably due to air leakage into the vacuum vessel. 
We describe quenching correction for a recoil with energy $E$ in terms of electronic $S_e(E)$ and nuclear $S_n(E)$
stopping powers, $q(E) \approx \left ( S_e(E) +  S_n(E) \cdot 0.3 \right ) / \left (  S_e(E) +  S_n(E) \right )$, where
factor 0.3 is chosen to be small and non-zero.
The quenching correction, $q$, which is about 25\% at 100~keV and 3\% at 900~keV~\cite{srim, Hitachi:2008kf},
is used in the calculation of light output for fluorine recoils.
The range of the recoil track is determined as the maximum distance between the pixels in the cluster.
Recoils are defined to have  at least 6 CCD bins that are not touching the boundary of the CCD view field.  
We require a ratio of principal moments of inertia greater than 5, which imposes a minimum energy cut of around 100~keV on nuclear recoils. 
Discharge events are removed with a cut on the maximum light  per event, and remaining spot-like events are removed
with the cut on the ration of principal moments.
A plot of energy versus range of simulated events and data is shown in Figure~\ref{fg::energy vs range}
for recoil candidates that have reconstructed energy above 100~keV.
The simulation is based on cross sections found in a nuclear scattering library~\cite{endf} and
fluorine ion propagation determined from the SRIM~\cite{srim}.

\begin{figure}[hb]
\center
\begin{tabular}{ll}
a) 935~keV & b) 550~keV \\
\includegraphics[width=6cm]{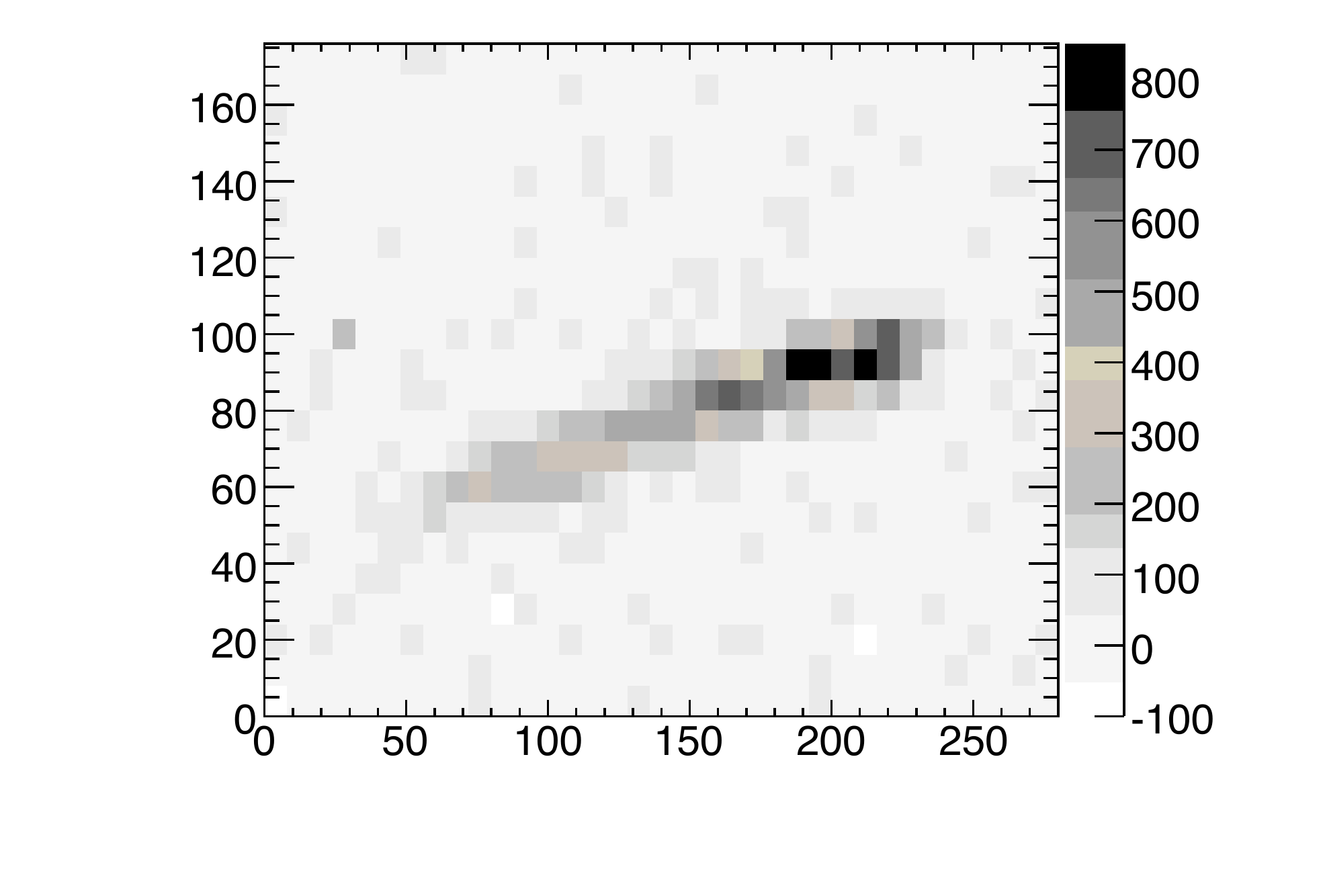} &
\includegraphics[width=6cm]{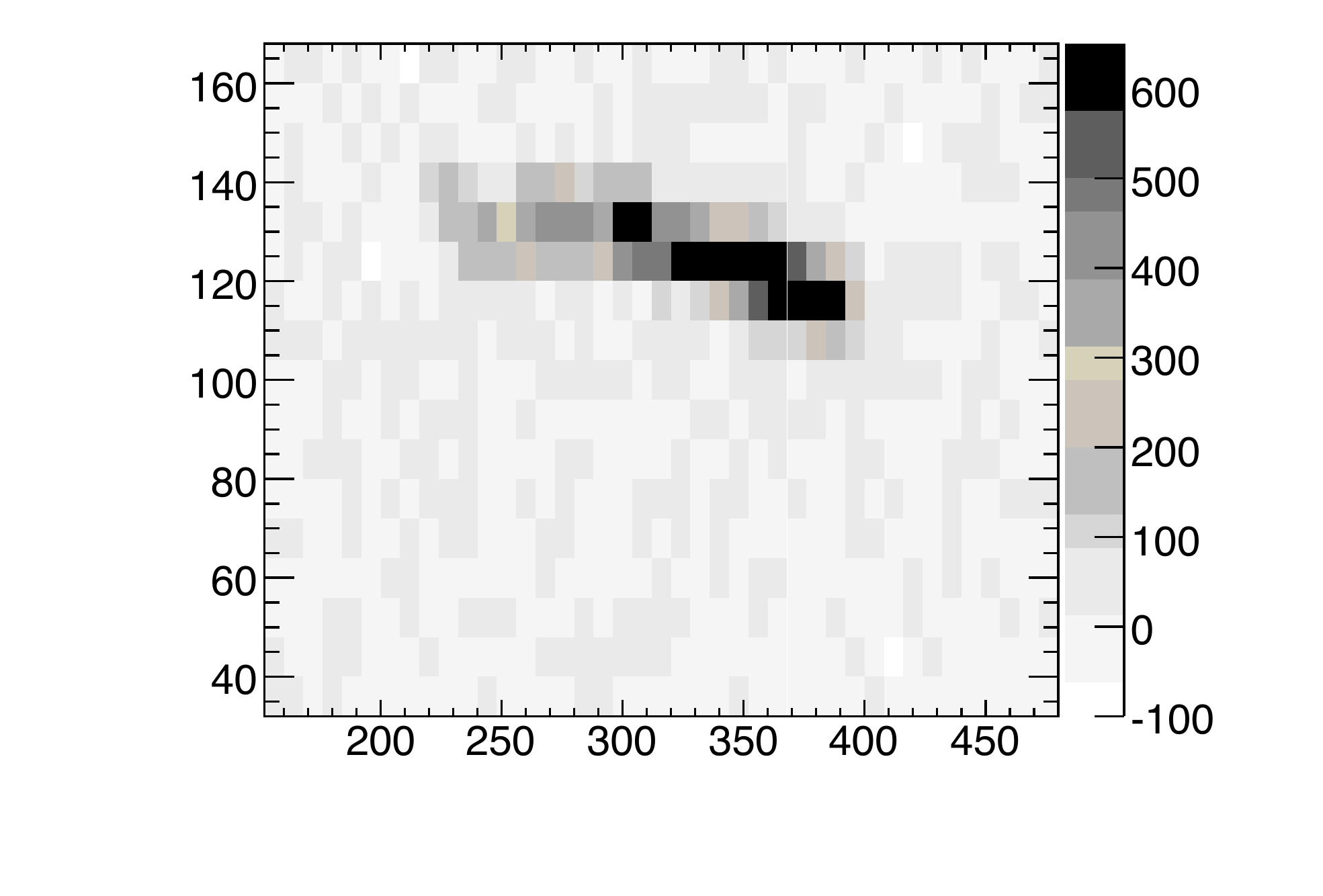} \\
a) 530~keV & b) 475~keV \\
\includegraphics[width=6cm]{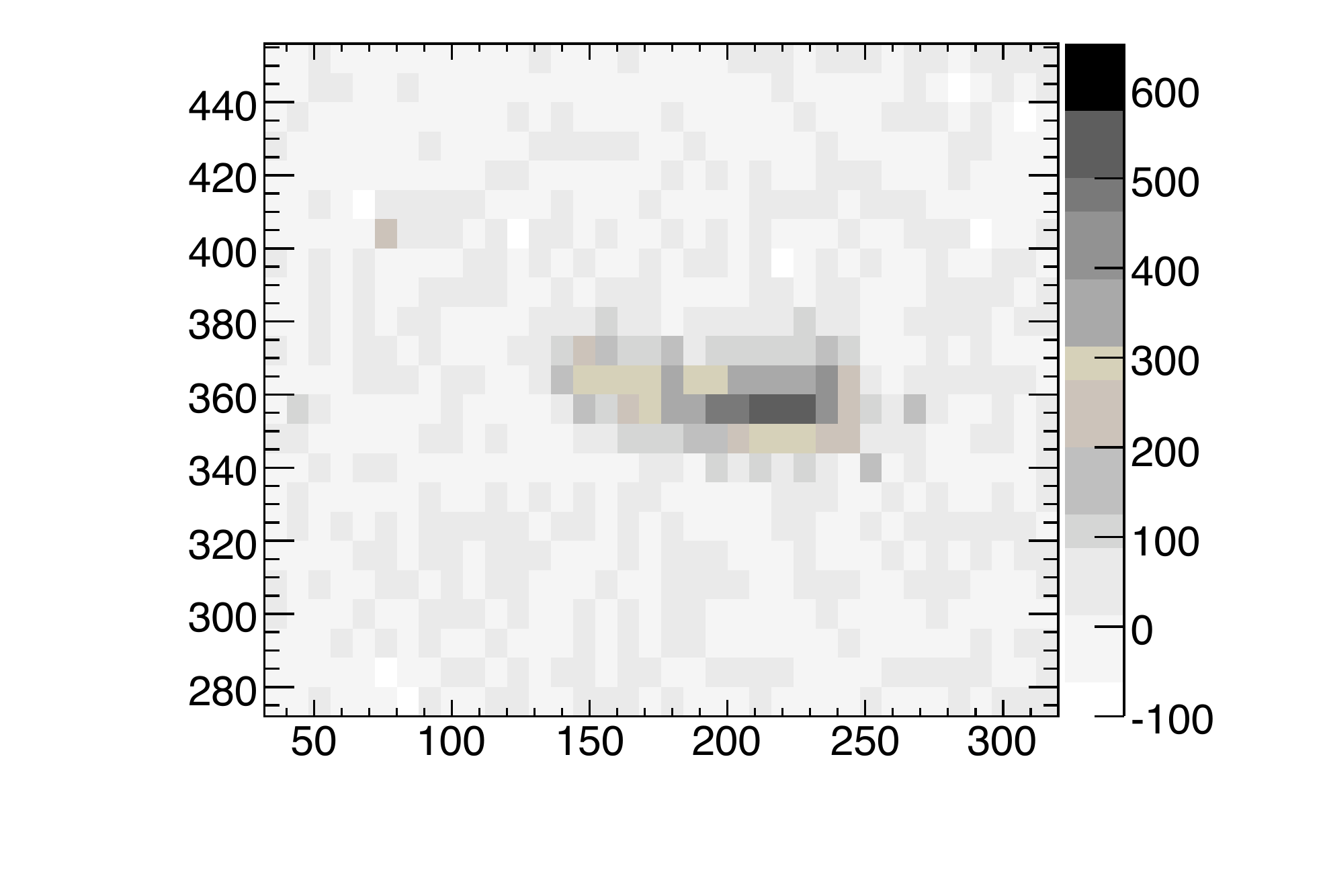} &
\includegraphics[width=6cm]{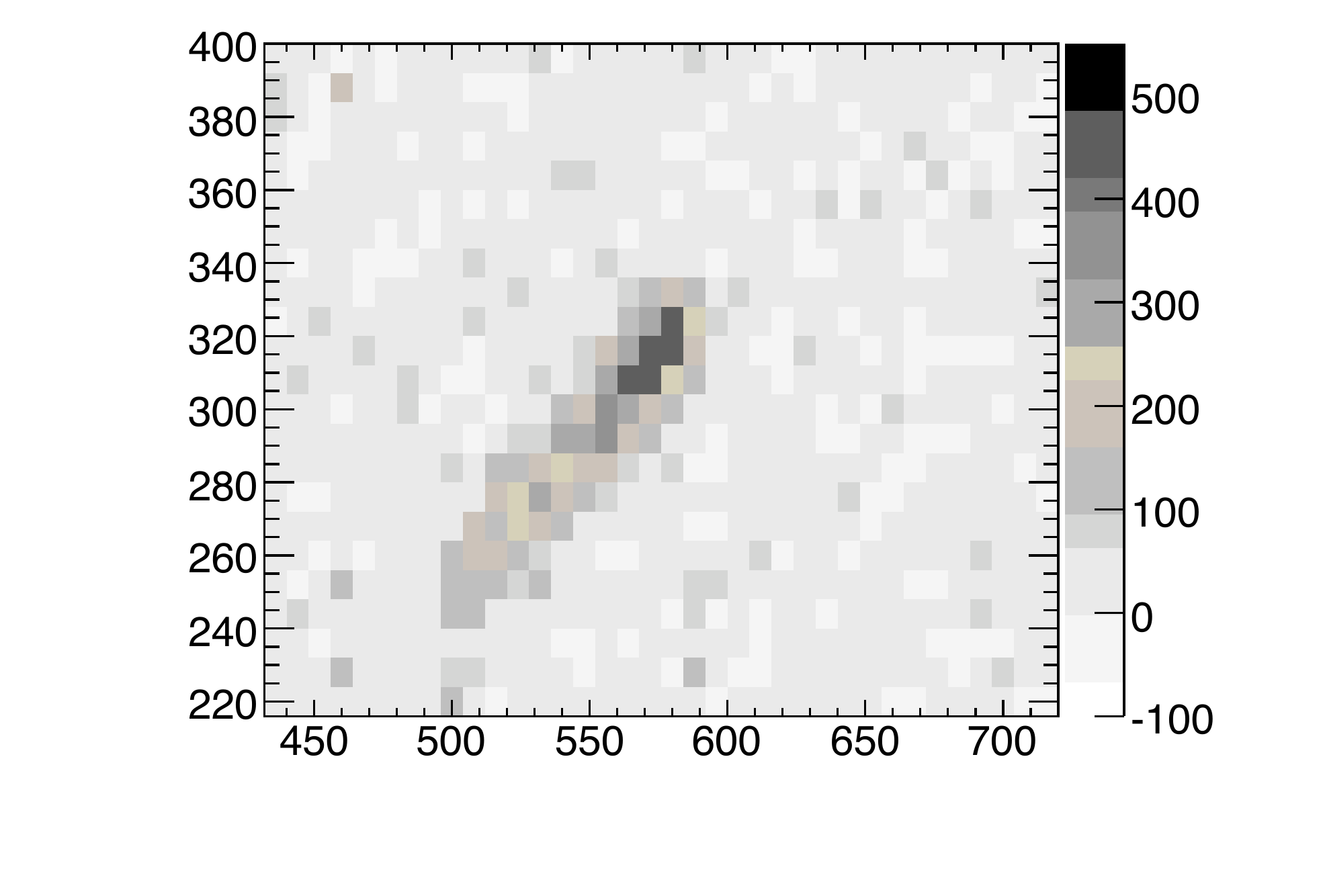} \\
c) 365~keV & d) 313~keV\\
\includegraphics[width=6cm]{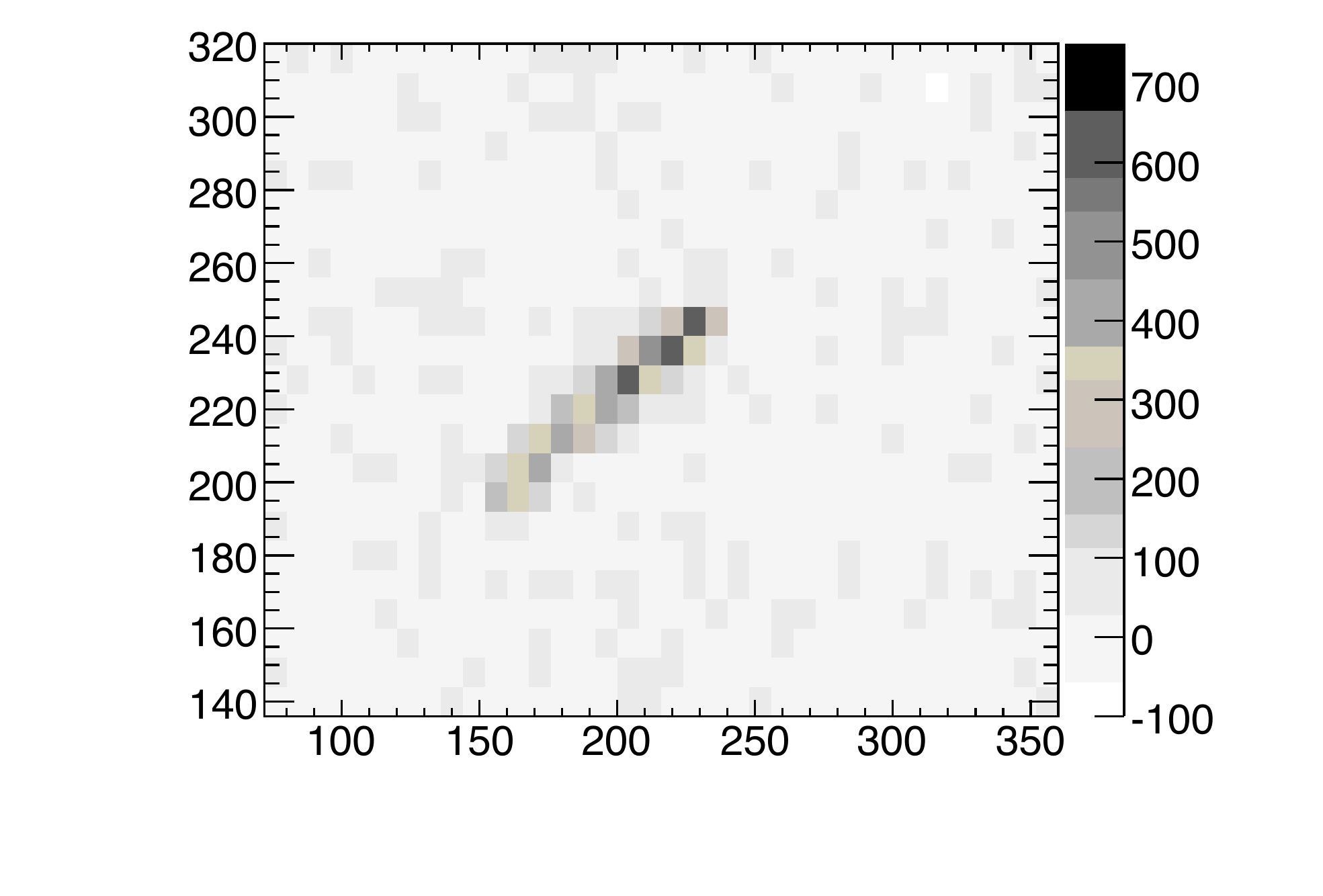} &
\includegraphics[width=6cm]{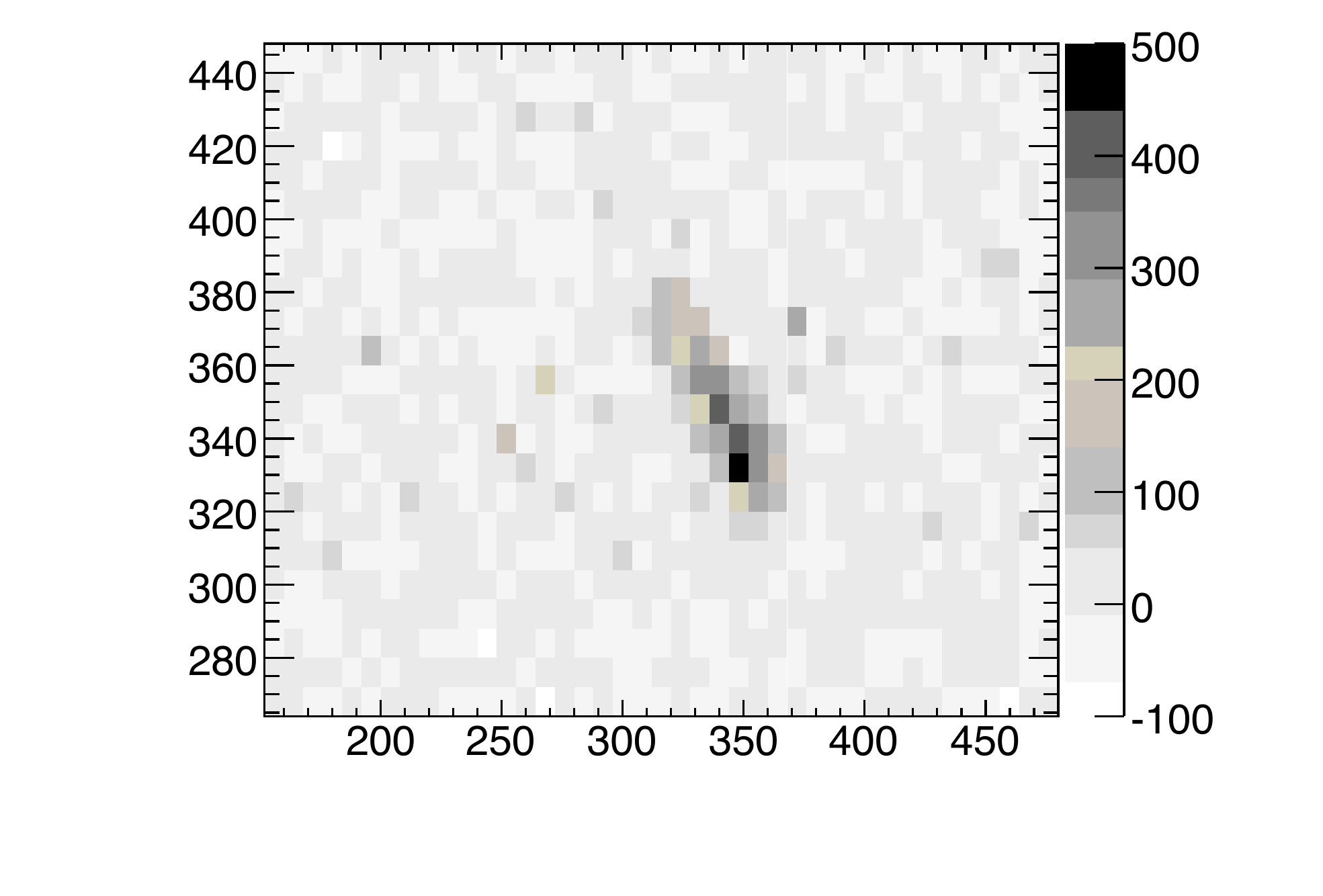} \\
\end{tabular}
\caption{Scintillation profile of nuclear recoil candidates in a $^{252}$Cf exposure 
at 75~Torr. Neutrons are incident from the right, along the $x$ axis.
\label{fg::recoil images}}
\end{figure}

We show images of recoil tracks with highest energies in Figure~\ref{fg::recoil images}.
In all images neutrons are incident from the right, along the $x$ axis.
The nuclear recoils also propagate from the right, and we observe a decreasing light intensity
as the recoils slow down in the \cf4 gas, as expected.
We quantify the scintillation asymmetry along the track using skewness, which is described previously in more detail~\cite{Dujmic:2007bd}. 
It is a dimensionless ratio of the third moment and the root-mean-square cubed of the light distribution along the track. 
Negative skewness is expected for recoils heading in the direction of the neutron beam.
The top-right plot in Figure~\ref{fg::energy vs range} shows the skewness distribution for tracks 
in data and simulation.

\begin{figure}[hb]
\center
\begin{tabular}{l}
\includegraphics[width=8cm]{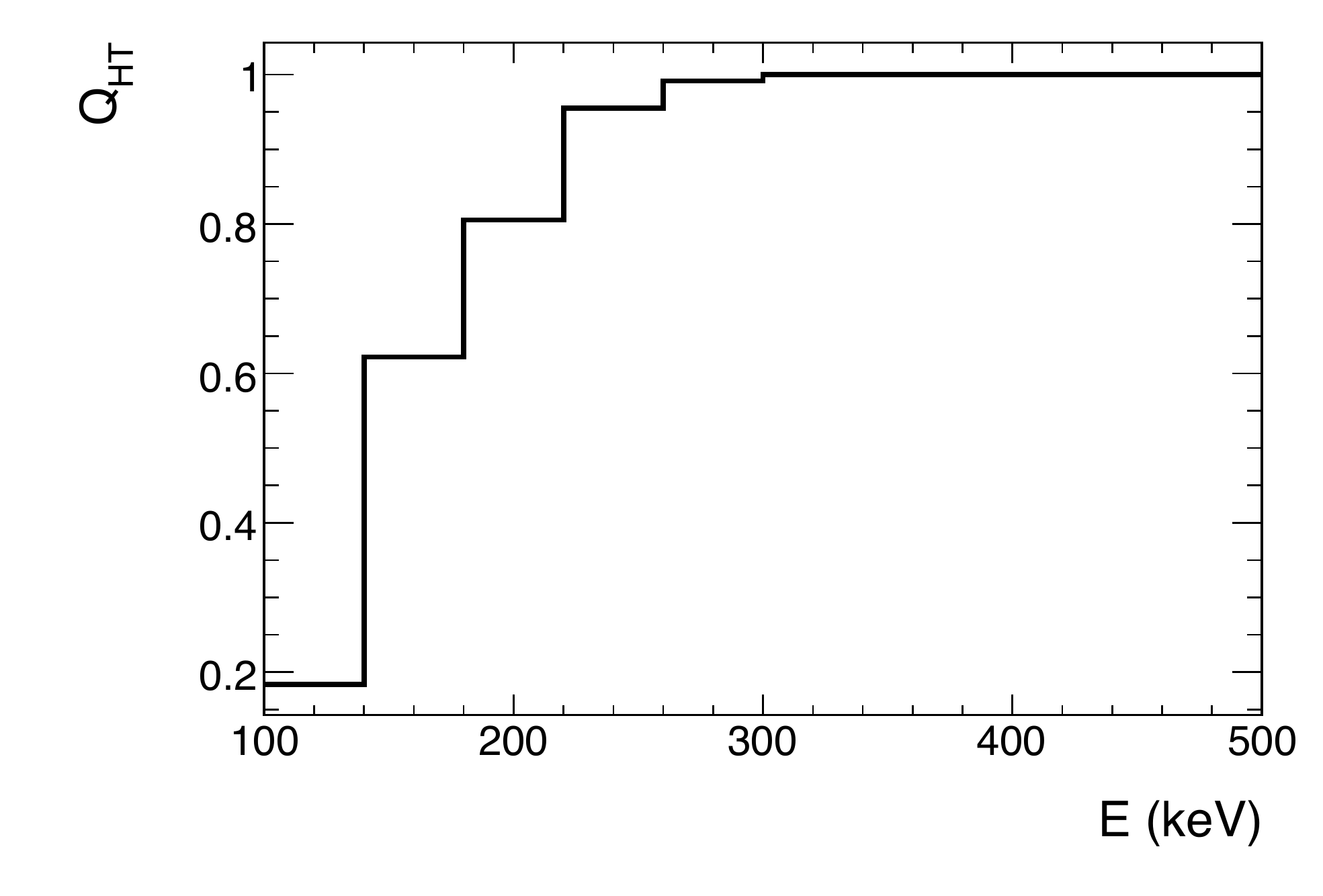} \\
\includegraphics[width=8cm]{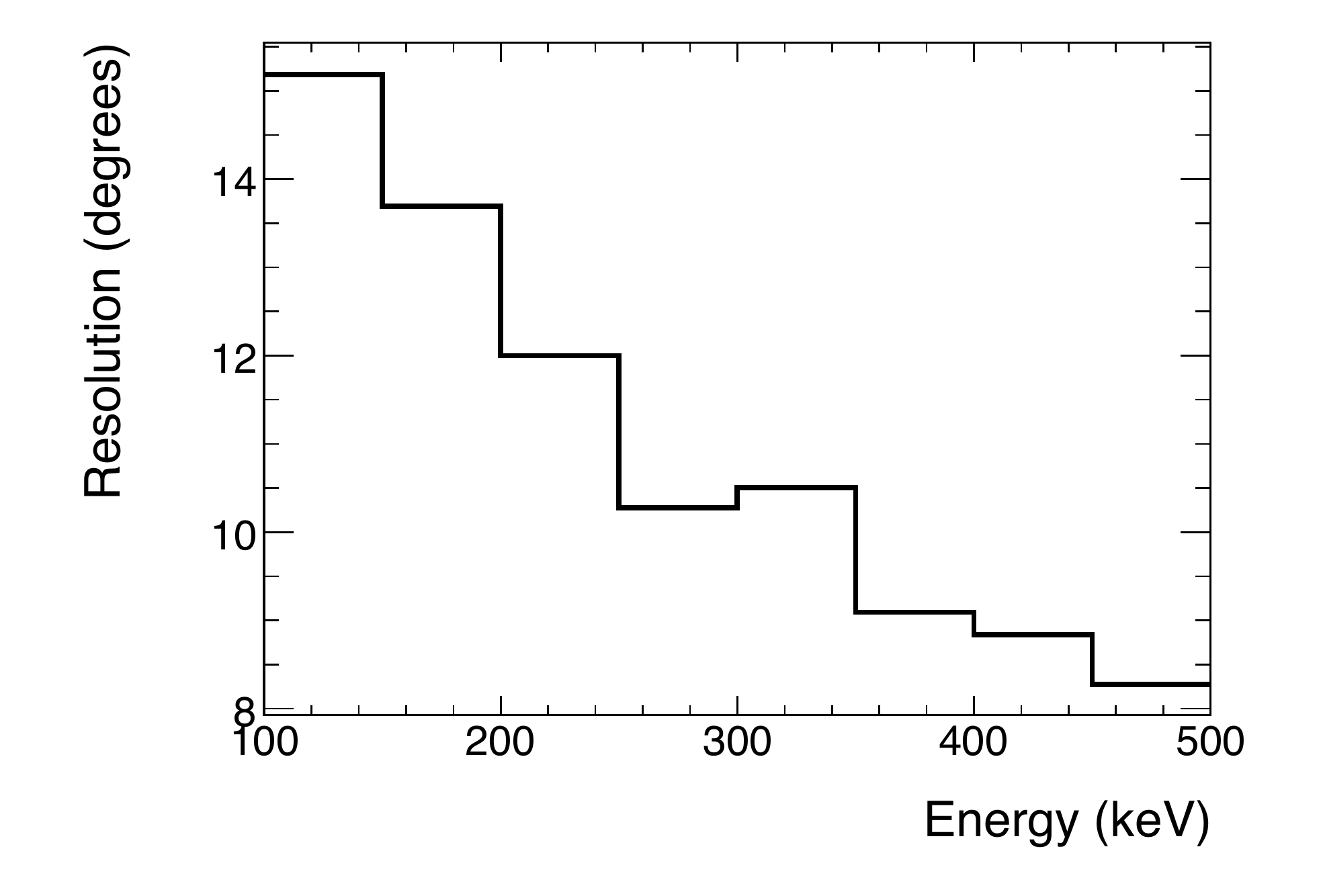}
\end{tabular}
\caption{Quality factor for the \headtail measurement vs. reconstructed energy (top),  
and 2D angular resolution  (bottom)  based on simulation for neutron scattering experiment.
\label{fg::Q and sigma angle}}
\end{figure}

The two pixels with maximum separation that are used in the range measurement are also used
in the calculation of the recoil angle. Figure~\ref{fg::Q and sigma angle} shows the resolution for the
recoil angle based on the simulation.
The sense of direction is determined from the \headtail (skewness) measurement.
We show for the first time, distributions for the signed cosine of the 2D recoil angle, $\cos\theta_{Recoil}$ vs. energy 
and signed cosine of the 2D recoil in Figure~\ref{fg::energy vs range}.

We define a quality factor, $Q_{HT}$ for the \headtail asymmetry at given recoil energy, $E_R$ as
\begin{equation}
        Q_{HT}(E_R) = \varepsilon(E_R) \cdot \left (1-2~\omega(E_R) \right )^2
\end{equation}
where the recoil reconstruction efficiency, $\varepsilon$ and the fraction of wrong \headtail assignments, $\omega$ are determined
from simulation. The $Q_{HT}$ is the effective fraction of reconstructed recoils with \headtail information, and
the error on the \headtail asymmetry scales as $1/\sqrt{Q_{HT}}$. 
Figure~\ref{fg::Q and sigma angle} shows the quality factor in the energy range that has been explored so far. 

\section{Conclusion and outlook}
\label{sec::conclusion}

\begin{figure}[hb]
\center
\includegraphics[width=10cm]{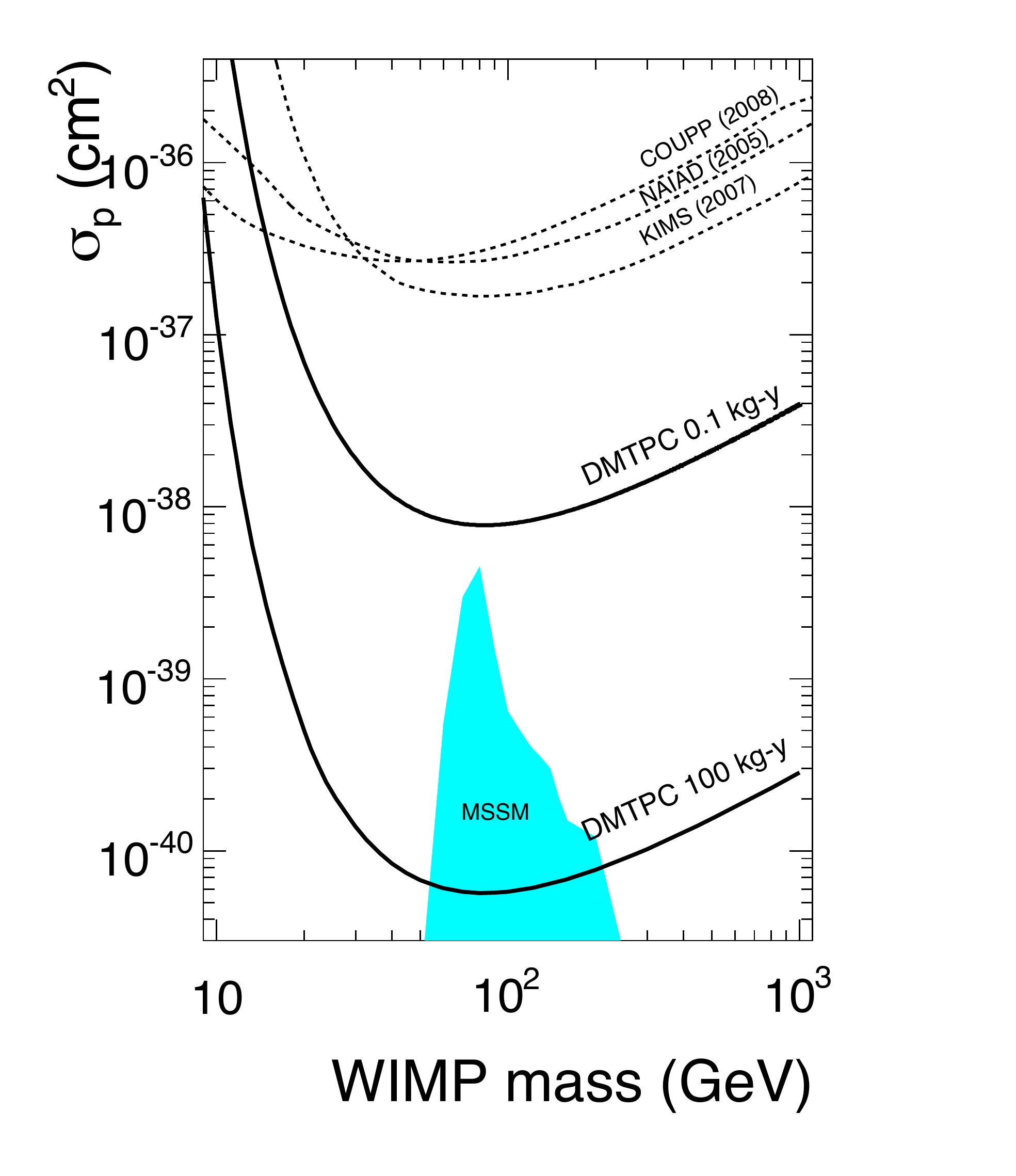} 
\caption{Estimated sensitivity (90\% C.L.) for spin-dependent WIMP scattering with a mesh detector.
We expect that MSSM models can be tested with approximately $100~\rm{kg}\cdot\rm{y}$ of target 
mass-exposure in an underground laboratory. 
\label{fg::WIMP}}
\end{figure}

Directional detection of dark matter  requires large  detector volumes with fine granularities.
TPC modules filled with low-pressure \cf4 gas can provide directionality and \headtail directional sense, 
in particular for spin-dependent dark matter searches.
We have demonstrated three possibilities for charge-amplification  that allow directional detection and  \headtail determination. 
In all three cases the gain is improved by more than an order of magnitude compared to the previous MWPC design~\cite{Dujmic:2007bd}.
This results in improvement in the \headtail discrimination due to the larger gain,
2D detection of recoil images, and  reduced pressure from 200 to 75~Torr,  allowing for longer recoil tracks.
%
%
%
We estimate the sensitivity for the WIMP detection using standard assumptions about the dark matter halo~\cite{Lewin:1995rx}.
Cross-section limits for spin-dependent WIMP scattering on proton are shown in Figure~\ref{fg::WIMP}. 
We can improve  current experimental limits~\cite{sd-exp} 
with approximately $0.1~\rm{kg}\cdot\rm{y}$ of \cf4 exposure, 
and test MSSM models~\cite{Ellis:2000jd} with approximately $100~\rm{kg}\cdot\rm{y}$. 
In both cases we assumed 
operation in an underground laboratory with a neutron-induced background rate of 
$0.01~\rm{events/(keV}\cdot\rm{kg}\cdot{y})$~\cite{Mei:2005gm}, uncorrelated with Cygnus direction, and a 50~keV recoil-energy threshold.
%
We plan to pursue these technologies and construct a cubic-meter module that will be a 
basic building block of a ton-scale detector.

\section{Acknowledgments} 

We wish to thank the Office of Environment, Health \& Safety at MIT
for supplying radioactive sources, the Laboratory for Nuclear Science MIT for technical support.
We thank Akira Hitachi and Jeff Filippini  for helpful discussions,
and Gus Zhang of Prochema for providing samples of ITO-mylar foils.
We acknowledge support by 
the Advanced Detector Research Program of the U.S. Department of Energy (contract number 6916448), 
the Reed Award Program, 
the Ferry Fund, 
the Pappalardo Fellowship program, 
the MIT Kavli Institute for Astrophysics and Space Research,
and the Physics Department at the Massachusetts Institute of Technology.


\begin{thebibliography}{00}

\bibitem{DMreview} 
  R.~J.~Gaitskell,
  Ann.\ Rev.\ Nucl.\ Part.\ Sci.\  {\bf 54}, 315 (2004).

\bibitem{Bernabei:2008yi}
  R.~Bernabei {\it et al.}  [DAMA Collaboration],
  arXiv:0804.2741 [astro-ph].

\bibitem{si-exp}
Recent limits on spin-independent WIMP interactions:
  J.~Angle {\it et al.}  [XENON Collaboration],
  Phys.\ Rev.\ Lett.\  {\bf 100}, 021303 (2008);
  Z.~Ahmed {\it et al.}  [CDMS Collaboration],
  arXiv:0802.3530 [astro-ph].


\bibitem{sd-exp}
Recent limits on spin-dependent WIMP interactions:
  G.~J.~Alner {\it et al.}  [UK Dark Matter Collaboration],
  Phys.\ Lett.\  B {\bf 616}, 17 (2005);
%
  E.~Behnke {\it et al.}  [COUPP Collaboration],
  Science {\bf 319}, 933 (2008);
%
  H.~S.~Lee. {\it et al.}  [KIMS Collaboration],
  Phys.\ Rev.\ Lett.\  {\bf 99}, 091301 (2007).



\bibitem{directionality}
  D.~N.~Spergel,
  Phys.\ Rev.\  D {\bf 37}, 1353 (1988);
%
   A.~M.~Green and B.~Morgan,
  Astropart.\ Phys.\  {\bf 27}, 142 (2007); 
%
  O.~Host and S.~H.~Hansen,
  JCAP {\bf 0706}, 016 (2007); 
%
  M.~S.~Alenazi and P.~Gondolo,
  Phys.\ Rev.\  D {\bf 77}, 043532 (2008).




\bibitem{DRIFT}
  S.~Burgos {\it et al.},
  Astropart.\ Phys.\  {\bf 28}, 409 (2007).
  P.~K.~Lightfoot, N.~J.~C.~Spooner, T.~B.~Lawson, S.~Aune and I.~Giomataris,
  Astropart.\ Phys.\  {\bf 27}, 490 (2007);
  D.~P.~Snowden-Ifft, C.~J.~Martoff and J.~M.~Burwell,
  Phys.\ Rev.\  D {\bf 61}, 101301 (2000);
  C.~J.~Martoff, D.~P.~Snowden-Ifft, T.~Ohnuki, N.~Spooner and M.~Lehner,
  Nucl.\ Instrum.\ Meth.\  A {\bf 440}, 355 (2000).



\bibitem{Miuchi:2007ga}
  K.~Miuchi {\it et al.},
  Phys.\ Lett.\  B {\bf 654}, 58 (2007);
  T.~Tanimori, H.~Kubo, K.~Miuchi, T.~Nagayoshi, R.~Orito, A.~Takada and A.~Takeda,
  Phys.\ Lett.\  B {\bf 578}, 241 (2004).

\bibitem{Santos:2007ga}
  D.~Santos, O.~Guillaudin, T.~Lamy, F.~Mayet and E.~Moulin,
  J.\ Phys.\ Conf.\ Ser.\  {\bf 65}, 012012 (2007).



\bibitem{Dujmic:2007bd}
  D.~Dujmic {\it et al.},
  Nucl.\ Instrum.\ Meth.\  A {\bf 584}, 327 (2008).
%


\bibitem{CF4 ionization} P. G. Datskos, J. G. Carter, and L. G. Christophorou, J. Appl. Phys. 71 (1982) 15.

\bibitem{Pansky:1994zh}
  A.~Pansky, A.~Breskin, A.~Buzulutskov, R.~Chechik, V.~Elkind and J.~Va'vra,
  Nucl.\ Instrum.\ Meth.\  A {\bf 354}, 262 (1995).


\bibitem{Kaboth:2008mi}
  A.~Kaboth {\it et al.},
  arXiv:0803.2195 [physics.ins-det].



\bibitem{Charpak:1997kd}
  G.~Charpak, R.~Bouclier, T.~Bressani, J.~Favier and C.~Zupancic,
  Nucl.\ Instrum.\ Meth.\  {\bf 62}, 262 (1968).


\bibitem{micropattern}
  F.~Angelini, R.~Bellazzini, A.~Brez, M.~M.~Massai, R.~Raffo, G.~Spandre and M.~A.~Spezziga,
  Nucl.\ Instrum.\ Meth.\  A {\bf 335}, 69 (1993);
%
  Y.~Giomataris, P.~Rebourgeard, J.~P.~Robert and G.~Charpak,
  Nucl.\ Instrum.\ Meth.\  A {\bf 376}, 29 (1996);
%
  F.~Sauli,
  Nucl.\ Instrum.\ Meth.\  A {\bf 386}, 531 (1997).




\bibitem{ccd readout}
	F.A.F. Fraga et al.. Nucl. Instr. and Meth. A 471 (2001), p. 125

\bibitem{Sharma:1998xw}
  A.~Sharma,
SLAC-JOURNAL-ICFA {\bf 16} (1998) 3.



\bibitem{srim} J.~F.~Ziegler, J.~P.~Biersack, U.~Littmark, 
  Pergamon Press, New York, 1985. The code is available online at www.SRIM.org.


\bibitem{Hitachi:2008kf}
  A.~Hitachi,
  arXiv:0804.1191.


\bibitem{endf} M.B. Chadwick {\it et al.}, Nuclear Data Sheets 107 (2006), 2931-3060


\bibitem{Mei:2005gm}
  D.~Mei and A.~Hime,
  Phys.\ Rev.\  D {\bf 73}, 053004 (2006).

\bibitem{Lewin:1995rx}
  J.~D.~Lewin and P.~F.~Smith,
  Astropart.\ Phys.\  {\bf 6}, 87 (1996).



\bibitem{Ellis:2000jd}
  J.~R.~Ellis, A.~Ferstl and K.~A.~Olive,
  Phys.\ Rev.\  D {\bf 63}, 065016 (2001).

\end{thebibliography}
\end{document}